\documentclass[aps,prd,reprint,nofootinbib,superscriptaddress,longbibliography]{revtex4-2}

\usepackage[T1]{fontenc}
\usepackage[utf8]{inputenc}
\usepackage{amsmath,amssymb,amsfonts,bm}
\usepackage{graphicx}
\usepackage{xcolor,ulem}
\usepackage{hyperref}
\usepackage{orcidlink}
\usepackage{booktabs}
\usepackage{multirow}
\usepackage{tikz}
\usetikzlibrary{positioning}
\usepackage{cleveref}
\usepackage{chngcntr}

\hypersetup{colorlinks=true,linkcolor=blue,citecolor=blue,urlcolor=blue}

\newcommand{\Sf}{S_{\rm f}}
\newcommand{\Sc}{S_{\rm c}}

\newcommand{\ee}{\mathrm{ee}}

\newcommand{\figplaceholder}[2]{%
\IfFileExists{#1}{\includegraphics[width=\linewidth]{#1}}{%
\fbox{\begin{minipage}[c][0.30\textheight][c]{0.92\linewidth}\centering
\texttt{#1}\\[0.5em]#2
\end{minipage}}}}

\begin{document}

\title{Renormalization-guided inverse blocking for lattice field generation: \\ construction and validation}

\author{Anna Hasenfratz\,\orcidlink{0000-0003-1813-2645}}
\email[Corresponding author: ]{anna.hasenfratz@colorado.edu}
\affiliation{Department of Physics, University of Colorado Boulder,
Boulder, Colorado 80309, USA}

\author{Ethan T.~Neil\,\orcidlink{0000-0002-4915-3951}}
\email[Corresponding author: ]{ethan.neil@colorado.edu}
\affiliation{Department of Physics, University of Colorado Boulder,
Boulder, Colorado 80309, USA}

\author{Letizia Parato\,\orcidlink{0000-0001-7500-6747}}
\email{letizia.parato@colorado.edu}
\affiliation{Department of Physics, University of Colorado Boulder,
Boulder, Colorado 80309, USA}

\author{Noah Schwartz}
\email{noah.schwartz-2@colorado.edu}
\affiliation{Department of Physics, University of Colorado Boulder,
Boulder, Colorado 80309, USA}
\date{\today}

\begin{abstract}
 We propose an algorithm for generating lattice field configurations based on
the approximate inversion of a renormalization-group blocking transformation.
We optimize the blocking transformation using a ``perfect blocking'' condition
so that the blocked lattice distribution is well approximated by a simple
coarse action.  The blocking is separated into an invertible smoothing
transformation followed by decimation.  Machine learning, in the form of a
conditional normalizing flow, is used to reconstruct the short-distance degrees
of freedom removed by the decimation.  A short fine-action rethermalization then
removes the residual mismatch.  Because the coarse ensemble supplies the
long-distance modes, the same blocking transformation and conditional flow can
be reused recursively on larger lattices, producing a cascade of configurations
from an initial small-volume ensemble.  We test the method in two-dimensional
$\phi^4$ theory with $\lambda=1$ at criticality and demonstrate stable cascade
upscaling from $16^2$ to $2048^2$ lattices on local computational resources.
Controlled rethermalization tests show that short-distance mismatches relax
rapidly, whereas a deliberately introduced mismatch in the relevant thermal
direction relaxes much more slowly.  The construction uses ingredients that
admit natural extensions to higher-dimensional systems and, ultimately, to
gauge and fermionic degrees of freedom. 
\end{abstract}

\maketitle

\section{Introduction}
\label{sec:introduction}

Markov-chain generation of lattice field configurations becomes increasingly expensive near a critical point, a phenomenon known as \emph{critical slowing down} \cite{Sokal:1997lke,Schaefer:2010hu,Finkenrath:2024ptc}.  While there are many interesting examples of theories with quantum critical points, one of the most important for high-energy physics is the Gaussian fixed point approached in the continuum limit of asymptotically free gauge theories such as QCD.  Simulations increasingly close to this limit suffer from the slow evolution of long-distance modes, most dramatically in the freezing of topological charge.

The origin of critical slowing down can be understood in terms of scale.  Most commonly used Monte Carlo algorithms evolve long-distance degrees of freedom much more slowly than short-distance fluctuations.  As the correlation length $\xi$ grows, increasingly large physical scales must be sampled, and the autocorrelation time rises rapidly.  Cluster algorithms such as the Wolff update \cite{Wolff:1989} can alleviate this problem in  spin systems, but no analogous general construction is known for gauge fields or fermions.

Normalizing-flow samplers offer a different approach by learning a probability density $q_\theta(\phi)$ that approximates the target Boltzmann distribution $p(\phi)\propto e^{-S(\phi)}$.  Because a flow provides both samples and their proposal probabilities, residual bias can be removed by reweighting or by a Metropolis correction step~\cite{Albergo:2019eim,Albergo:2021vyo}. Normalizing-flow constructions have also been extended to gauge and
fermionic systems~\cite{Boyda:2020hsi,Albergo:2021bna,Abbott:2022zhs,Abbott:2022zsh}. 

A perfectly trained flow would generate statistically independent samples directly from the target distribution.  In practice, however, whole-volume flows also encounter severe volume scaling \cite{DelDebbio:2021qwf,Komijani:2023fzy}.  A flow trained from an unstructured base distribution must learn long-distance critical fluctuations and short-distance structure simultaneously. In addition, at approximately fixed accuracy per degree of freedom, residual errors accumulate extensively with the volume, causing the acceptance of global independence proposals to deteriorate rapidly, requiring increasingly accurate and costly training. 

We instead approach the problem by reversing a renormalization-group (RG) blocking transformation.  Consider a target fine-lattice action $S_f$.  Blocking produces a coarse field distributed according to the blocked action
\begin{equation}
    S_b = B[S_f] .
\end{equation}
The blocked ensemble retains the long-distance physics of the fine theory while short-distance degrees of freedom have been removed.  If this coarse distribution can be sampled efficiently, the difficult infrared part of the field-generation problem is already solved: reconstructing the fine ensemble requires adding back only the information discarded by blocking.  
We refer to this coarse-to-fine construction as \emph{upscaling}.  Related ideas appear in multiscale equilibration and prolongation methods
\cite{Endres:2015yca}, as well as in inverse-blocking and super-resolution
approaches in statistical systems and lattice field theory
\cite{PhysRevLett.89.275701,Bachtis:2021eww,Wang:2023exq,Marchand:2022fxp,Bauer:2024byr,Zhu:2025pmw,Masuki:2025pfx,Tan:2026ibs,Masuki:2026qfw,Singha:2026slc}.

The central obstacle is that $S_b$ is generally unknown.  RG blocking generates additional couplings, so even a simple fine action can map into a complicated action in an effectively infinite-dimensional coupling space.  The closely related perfect-action program addresses this difficulty by constructing actions on or near the renormalized trajectory of a given block transformation, for which cutoff effects and RG evolution are particularly simple \cite{Hasenfratz:1993sp,DeGrand:1995ji,DeGrand:1995jk,Holland:2024muu,Holland:2025fsa}.  Here we take the complementary approach of optimizing the RG transformation itself so that the blocked distribution is well approximated by a simple coarse action $S_c$,
\begin{equation}
    e^{-S_c} \simeq e^{-B[S_f]}.
\end{equation}
 We refer to this construction as \emph{perfect blocking}.  The freedom to optimize the RG transformation itself has been explored previously in Refs.~\cite{Swendsen:1984vu,Hasenfratz:1984hx,Tu:2018dws}.  When the matching is successful, configurations generated directly with $S_c$ can replace explicitly blocked fine configurations and supply the infrared input for the inverse transformation.  In general the matched coarse and fine actions have different couplings; they coincide only at an RG fixed point.

The resulting construction combines approximate perfect blocking with conditional reconstruction of the degrees of freedom removed by blocking.  The upscaled configurations carry the correct long-distance structure.  We repair any short-distance mismatch by rethermalization using a local Monte Carlo algorithm such as hybrid Monte Carlo (HMC) \cite{Duane:1987de}.
Here we develop the construction and its diagnostics in detail, with particular emphasis on the origin and control of rethermalization errors.  Controlled tests show a pronounced separation of scales: even large microscopic mismatches can be corrected rapidly when the inherited infrared distribution is correct, whereas deliberately mistuning the relevant (thermal) coupling produces relaxation of long-distance observables over a much longer Monte Carlo time scale.  This distinction provides a direct test of the central premise of the method---that the inverse-RG construction should supply the slow infrared degrees of freedom before fine-lattice evolution is applied.

We demonstrate the method in the two-dimensional one-component scalar $\phi^4$ theory,
\begin{equation}
  S[\phi]= -2\kappa\sum_{x,\mu}\phi_x\phi_{x+\hat\mu}
  +\sum_x \left[\phi_x^2+\lambda(\phi_x^2-1)^2\right] .
  \label{eq:phi4-action}
\end{equation}
For the main numerical study we set $\lambda=1$ and work at the critical point,
$\kappa\simeq\kappa_{\rm cr}=0.340301$ \cite{Bosetti:2015lsa,Kaupuzs:2025phi4}.  The blocking transformation and conditional model are determined using only small-volume ensembles. Working at criticality, the perfect blocking gives $S_f(\kappa_c) \approx S_c(\kappa_c)$, allowing us to train once and then apply the upscaling repeatedly to generate configurations up to very large volumes; we demonstrate the method up to $L=2048$. Detailed comparisons with directly generated (``native'') Monte Carlo ensembles, RG-invariant observables, finite-size scaling, and controlled rethermalization studies are used below to test both the physical fidelity of the generated ensembles and the assumptions underlying the algorithm.

\section{Overview of the method}
\label{sec:method_overview}

\begin{figure*}[t]
\centering   
    \usetikzlibrary{positioning,calc,arrows.meta}
    \colorlet{diagramblue}{blue}
    \colorlet{trainpurple}{purple}
    \resizebox{\linewidth}{!}{%
    \begin{tikzpicture}[
        >=Latex,
        box/.style={
            draw=black,
            rounded corners=1.5mm,
            line width=0.9pt,
            minimum height=0.95cm,
            inner xsep=6pt,
            inner ysep=4pt,
            align=center
        },
        arrow/.style={
            -{Latex[length=2.0mm,width=2.2mm]},
            line width=1.0pt
        },
        train/.style={
            trainpurple,
            dashed,
            line width=0.8pt
        },
        title/.style={
            font=\large
        }
        ]
    \node[box] (fine) at (0,0) {
        \large
        \textcolor{diagramblue}{$\{\phi_f\}\sim e^{-S_f(\kappa)}$}
        ~
        \textcolor{black}{$\big|\,L^d$}
    };
    \node[title, above=1mm of fine] {Fine field};
    \node[box] (smooth) at (5.6,0) {
        \large
        \textcolor{diagramblue}{$\psi=K\phi_f$}
        ~
        \textcolor{black}{$\big|\,L^d$}
    };
    \node[title, above=1mm of smooth] {Smoothed field};
    \node[box] (blocked) at (11.2,0) {
        \large
        \textcolor{diagramblue}{$\phi_b=D_2\psi$}
        ~
        \textcolor{black}{$\big|\,\left(L/2\right)^d$}
    };
    \node[title, above=1mm of blocked] {Blocked field};
    \node[box] (coarse) at (17.2,0) {
        \large
        \textcolor{diagramblue}{
            $\{\phi_b\}\approx\{\phi_c\}\sim e^{-S_c}$
        }
        ~
        \textcolor{black}{$\big|\,\left(L/2\right)^d$}
    };
    \node[title, above=1mm of coarse] {Matched coarse field};
    \draw[arrow]
        (fine.east) --
        node[above, text=trainpurple, font=\large] {$K$}
        (smooth.west);
    \draw[arrow]
        (smooth.east) --
        node[above, text=diagramblue, font=\large] {$D_2$}
        (blocked.west);
    \draw[arrow]
        (blocked.east) --
        (coarse.west);
    \node[box] (finecoarse) at (0,-3.4) {
        \large
        \textcolor{diagramblue}{$\phi\to\phi_c$}
        ~
        \textcolor{black}{$\big|\,L\to2L$}
    };
    \node[title, above=1mm of finecoarse] {Fine becomes coarse};
    \node[box] (retherm) at (5.6,-3.4) {
        \large
        \textcolor{diagramblue}{
            $\phi_{\rm prop}\to\phi\sim e^{-S_f(\kappa)}$
        }
        ~
        \textcolor{black}{$\big|\,L^d$}
    };
    \node[title, align=center, above=1mm of retherm]
        {Rethermalize\\[-1mm](HMC or cluster)};
    \node[box] (inverse) at (11.2,-3.4) {
        \large
        \textcolor{diagramblue}{
            $\phi_{\rm prop}=K^{-1}\psi_{\rm prop}$
        }
        ~
        \textcolor{black}{$\big|\,L^d$}
    };
    \node[title, align=center, above=1mm of inverse]
         {Proposed fine field};    
    \node[box] (proposal) at (17.2,-3.4) {
        \large
        \textcolor{diagramblue}{
            $\psi_{\rm prop}=(\phi_c,d)$
        }
        ~
        \textcolor{black}{$\big|\,L^d$}
    };
    \draw[arrow]
        (coarse.south)
        --
        node[pos=0.4, left, text=trainpurple, font=\large] {$\{d\}\sim q_\theta(\cdot\mid\phi_c)$} 
        (proposal.north);
    \draw[arrow]
        (proposal.west)
        --
        node[above, text=diagramblue, font=\large] {$K^{-1}$}
        (inverse.east);
    \draw[arrow]
        (inverse.west) --
        (retherm.east);
    \draw[arrow]
        (retherm.west) --
        (finecoarse.east);
    \draw[train]
        ($(fine.north east)+(0.9,0.10)$)
        --
        ($(fine.north east)+(0.9,1.00)$)
        --
        ($(coarse.north)+(0,1.00)$)
        --
        ($(coarse.north)+(0,0.60)$);
    \node[
        text=trainpurple
    ] at (9.0,1.8)
    {
        $K$ trained such that
        $\{\phi_b\}\approx\{\phi_c\}$
    };
    \draw[train]
        ($(smooth.south)+(0,-0.15)$)
        --
        ($(smooth.south)+(0,-1.)$)
        --
        ($(coarse.south)+(-3.2,-1.)$);
    \node[
        text=trainpurple
    ] at (9.6,-1.2)
    {
        Conditional flow trained on $\{\psi\}$
    };
    \draw[arrow]
        ($(finecoarse.south)+(0,-0.20)$)
        --
        ($(finecoarse.south)+(0,-0.85)$)
        --
        ($(proposal.south)+(0,-0.85)$)
        --
        ($(proposal.south)+(0,-0.20)$);
    
    \node[font=\large] at (8.6,-5.15) {
        Cascade sequence:
        $L\to2L\to4L\to8L\to\cdots$
    };
    \end{tikzpicture}%
    }
    \caption{
    Schematic overview of the renormalization-guided flow construction.
    The forward RG map of the top row defines the matched coarse distribution.  The inverse construction in the middle row generates detail fields conditionally on the coarse field and reconstructs the fine field with $K^{-1}$. The last row shows the fine-action rethermalization used to remove residual mismatch, followed by the recursive cascade to increasingly large volumes.
    }
    \label{fig:method_cartoon}
\end{figure*}
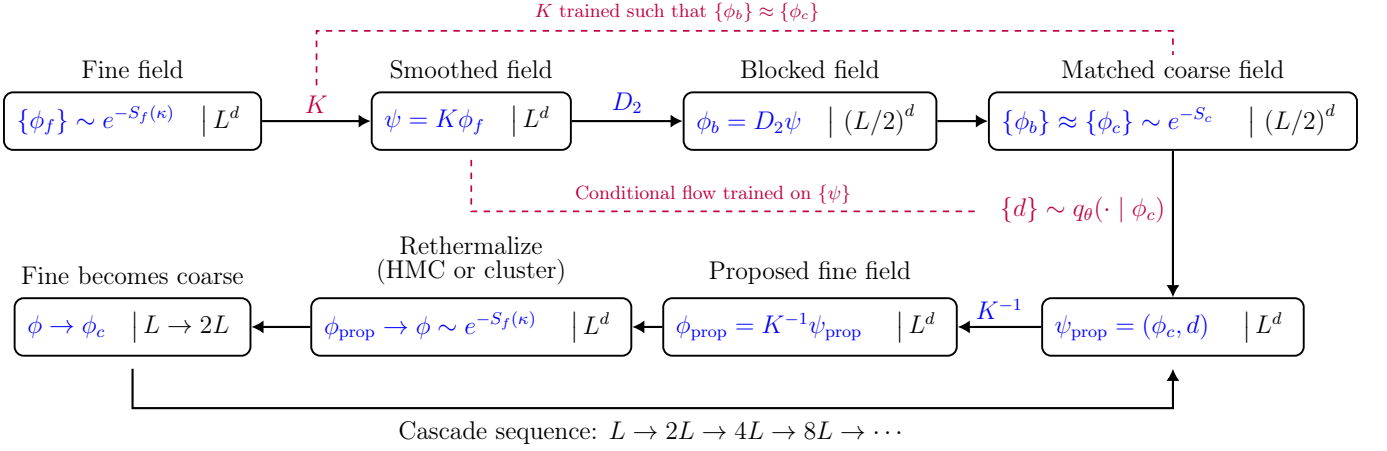

Our algorithm separates the infrared and ultraviolet parts of the field-generation problem. The overall construction is summarized in Fig.~\ref{fig:method_cartoon}. 

Let $S_f(\kappa_f)$ denote the target fine-lattice action. 
A scale-$s$ real-space RG blocking transformation $B$ maps this action to a blocked action
\begin{equation}
    S_b = B[S_f(\kappa_f)] .
\end{equation}

In general $S_b$ is not known explicitly.  We therefore introduce a tractable coarse action
$S_c(\kappa_c)$ whose ensemble approximates the blocked distribution,
\begin{equation}
    S_b \simeq S_c(\kappa_c).
\end{equation}
The coarse and fine couplings are matched so that the long-distance physics agrees under the RG transformation. 
In particular, the correlation lengths satisfy
\begin{equation}
    \xi_c(\kappa_c) \simeq \frac{\xi_f(\kappa_f)}{s}.
\end{equation}
At an RG fixed point the matched couplings coincide,
\begin{equation}
    \kappa_c=\kappa_f=\kappa_\star ,
\end{equation}
which is the special case used in our critical-point numerical study.

We implement $B$ in two steps.  First we apply an invertible smoothing transformation (a linear convolution)
\begin{equation}
    \psi = K\phi ,
\end{equation}
followed by a decimation $D_s$ that retains one residue class,
\begin{equation}
    \phi_b = D_s\psi = D_s K\phi .
\end{equation}
For $s=2$ in two dimensions, the retained variables are the even-even sites of $\psi$. 
The remaining residue classes are denoted by
\begin{equation}
    d_{01},\ d_{10},\ d_{11},
\end{equation}
so that
\begin{equation}
    \psi=(\phi_b,d_{01},d_{10},d_{11}).
\end{equation}
The full blocking map $D_sK$ is non-invertible because of the decimation, but the change of variables
$\phi\leftrightarrow\psi$ is chosen to be invertible.

If $\phi_c$ is drawn from a coarse distribution matched to $S_b$, it can be used in place of the blocked field $\phi_b$. 
The remaining task is then to reconstruct the missing detail fields conditioned on $\phi_c$. 
We use a conditional normalizing flow to model
\begin{align}
    d_{01} &\sim q_{\theta,01}(d_{01}\mid\phi_c),\\
    d_{10} &\sim q_{\theta,10}(d_{10}\mid\phi_c,d_{01}),\\
    d_{11} &\sim q_{\theta,11}(d_{11}\mid\phi_c,d_{01},d_{10}),
\end{align}
after which the proposed fine field is reconstructed as
\begin{equation}
    \phi_{\rm prop}=K^{-1}(\phi_c,d_{01},d_{10},d_{11}).
\end{equation}

The normalizing flow is  not used as a standalone exact sampler. Instead, it provides an RG-informed initialization in which the long-distance modes are supplied by the matched coarse ensemble, while the conditional model reconstructs the missing short-distance degrees of freedom.

The conditional flow need not reproduce the exact fine distribution. 
Instead, the proposed configuration is  evolved with a Markov update whose stationary distribution is
$e^{-S_f}$; we refer to this finite fine-action evolution as rethermalization. 
If the matched coarse ensemble has supplied the correct infrared distribution, the remaining mismatch is  dominated by short-distance structure and can be removed with comparatively few fine-action updates. 
Any residual finite-rethermalization bias can be tested by varying the number of update steps.

After rethermalization, the resulting fine configuration can itself be used as the coarse input for another inverse-blocking step. 
Repeating the construction gives the cascade
\begin{equation}
    L\rightarrow sL\rightarrow s^2L\rightarrow\cdots .
\end{equation}
Independent root configurations generate independent cascades, while configurations at different volumes within a single cascade remain strongly correlated because they share the same inherited infrared structure.

The construction therefore has three essential components, which we describe in turn below: the approximate perfect-blocking kernel $K$, the conditional upscaling flow, and the fine-action rethermalization.

\section{Perfect blocking}
\label{sec:perfect-why}

The central requirement of inverse blocking is that the coarse ensemble reproduce the probability distribution obtained by RG blocking the fine target ensemble, $S_c \simeq S_b$.  If this matching fails, the coarse field carries incorrect infrared information that cannot be repaired efficiently by a conditional model designed to reconstruct only the missing detail variables.

\subsection{Path-integral picture}

Formally, the path integral makes the perfect-blocking construction precise.
We begin with the partition function of the fine theory,
\begin{equation}
Z_f=\int D\phi\,e^{-S_f[\phi]}.
\end{equation}
Under the invertible linear transformation $\psi=K\phi$,
\begin{equation}
D\phi = |\det K|^{-1}D\psi ,
\end{equation}
and hence
\begin{equation}
Z_f
 = |\det K|^{-1}\int D\psi\,
   e^{-\widetilde S[\psi]},
\qquad
\widetilde S[\psi]\equiv S_f[K^{-1}\psi].
\end{equation}
We then partition the smoothed field into the retained blocked field
$\phi_b$ and the three detail fields, collectively denoted by $d$,
and integrate over the latter:
\begin{align}
Z_f
 &= |\det K|^{-1}
    \int D\phi_b\,Dd\,
    e^{-\widetilde S[\phi_b,d]}
    \notag\\
 &= \mathcal N
    \int D\phi_b\,e^{-S_b[\phi_b]} .
\label{eq:def_Sb}
\end{align}
Equation~\eqref{eq:def_Sb} defines the blocked action $S_b[\phi_b]$,
up to an additive constant.  Although the transformed action
$\widetilde S$ is determined by $S_f$ and $K$, integrating out the detail
fields generally generates many additional interactions; thus, $S_b$ is a
complicated functional of $\phi_b$ and may contain an effectively infinite
set of couplings.

If the exact blocked action $S_b$ could be sampled, configurations drawn
from $e^{-S_b}$ would have exactly the same probability distribution as the
blocked fields obtained from the fine target ensemble.  They would therefore
provide the correct infrared input for the inverse transformation, leaving
only the conditional reconstruction of the discarded detail variables.

In practice, the exact blocked action is not available in a tractable form, and we do not require
that blocking generates no additional operators.  Rather, we choose a simple
family of coarse actions
\begin{equation}
S_c[\phi]=S_{\rm lat}[\phi;\kappa_c,\lambda_c]
\end{equation}
and optimize the blocking transformation so that the probability distribution
generated by $S_c$ approximates the exact blocked distribution as closely as
possible,
\begin{equation}
e^{-S_c} \simeq e^{-B[S_f]} .
\label{eq:perfect-condition-general}
\end{equation}
The coarse action $S_c$ then serves as a practical sampler for the blocked
distribution.  We refer to a transformation satisfying this condition with good
accuracy as an \emph{approximate perfect blocking}.

For the numerical study below, we fix $\lambda_c=\lambda_f=1$.  At the
critical point, the matched relevant couplings coincide,
$\kappa_c=\kappa_f=\kappa_{\rm cr}$, so the practical matching condition becomes
\begin{equation}
    e^{-B[S_f(\kappa_{\rm cr})](\phi)}
    \simeq
    e^{-S_c(\kappa_{\rm cr})(\phi)} ,
\label{eq:perfect-condition}
\end{equation}
as tested through expectation values and distributional overlap of blocked and
directly generated coarse observables.  When this condition is satisfied, a
configuration drawn from $S_c$ is distributed approximately as an RG-blocked
configuration of the fine theory.  It therefore supplies the correct
long-distance input to the conditional flow, which need only reconstruct the
discarded ultraviolet details.

The condition in Eq.~\eqref{eq:perfect-condition} is practical rather than exact.  We require sufficient overlap between the directly generated coarse ensemble and the ensemble obtained by blocking fine configurations so that the residual mismatch after upscaling can be removed with a short fine-action evolution.  The quality of this matching is therefore crucial: residual short-distance mismatch can relax rapidly, whereas a mismatch in the relevant thermal direction changes the inherited infrared distribution and leads to much slower equilibration of long-distance observables.  We test this separation of scales explicitly in Sec.~\ref{sec:rethermalization}.

\subsection{Approximate perfect blocking kernel}
\label{sec:perfect-kernel}

The smoothing kernel $K$ is taken to be linear and translationally invariant and is optimized so that the retained field
\begin{equation}
    \psi_{\ee}\equiv D_{\ee}K\phi,
\end{equation}
obtained by retaining the even-even sites from smoothed fine configurations $\psi$, matches the distribution of directly generated coarse configurations as closely as possible.  Smoothing acts as a convolution on the field, i.e.
\begin{equation}
\psi(\vec{x}) = (K\phi)(\vec{x}) = \sum_{\vec{r}} K(\vec{r})\, \phi(\vec{x} + \vec{r})
\end{equation}
with periodic boundary conditions on the field in all directions.  The symmetry $K(-\vec{r}) = K(\vec{r})$ has been used to rewrite a standard convolution in a form that more closely resembles a weighted average of field values.

We use a $D_4$-symmetric $7\times7$ blocking kernel with ten orbit
coefficients, corresponding to nine independent fit parameters once the
overall normalization is fixed.  The production kernel was optimized by
comparing directly generated $L=16$ configurations with $L=32$
configurations blocked to $L=16$, at $\lambda=1$ and
$\kappa=0.340301$.  Its normalization was fixed to
\begin{equation}
    \sum_r K(r)=2^{\eta/2},\qquad \eta=0.25,
\end{equation}
and this normalization factor is already included in the coefficients
quoted below. \footnote{The normalization can alternatively be included among the
parameters determined by the perfect-blocking optimization.  Allowing it
to float gives a value corresponding to
$\eta_{\rm pred}\simeq0.251(1)$, consistent with the exact Ising value
$\eta=1/4$.  The parenthetical uncertainty denotes the spread among
optimizer starts rather than a statistical confidence interval.}

 The fitting objective minimizes a covariance-weighted mismatch over a set of
local and long-distance observables. Schematically,
  \begin{equation}
  \chi^2_K = \sum_{a,b}\left(\langle O_a\rangle_{B\Sf}
  -\langle O_a\rangle_{\Sc}\right)
  C^{-1}_{ab}
  \left(\langle O_b\rangle_{B\Sf}
  -\langle O_b\rangle_{\Sc}\right),
  \label{eq:kernel-objective}
\end{equation}
We have also experimented with adding a locality penalty term to the cost function, of the form $\Delta \chi_K^2 = \mu\ f(K^{-1})$, where $\mu$ is an adjustable hyperparameter and $f(K^{-1})$ measures the fraction of the total weight of the inverse kernel $K^{-1}$ which lies outside of an $5 \times 5$ box.  In some cases the optimal kernel was found to be at $\mu > 0$, but for the final kernel tuning described here the optimum is at $\mu=0$ and the locality penalty need not be imposed.

The observables used to optimize and validate the blocking transformation,
and later to diagnose the conditional flow and rethermalization, are
defined in \cref{app:operators}.

The fitted operator set is
    \begin{align}
         {\cal O}_{\rm fit}=
   \left\{
\begin{aligned}
   &\phi^2,\phi^4,\phi^6,\mathrm{NN},2\mathrm{NN},\mathrm{diag},
   m^2, \nonumber \\
   &G_{21},G_{22},G_{30},G_{31}
   \end{aligned}
   \right\}\, ,  
    \end{align}
  where $G_{ab}$ denotes the two-point function at the corresponding
  lattice separation.

The operators $m^4$, the local kurtosis
$\langle\phi^4\rangle/\langle\phi^2\rangle^2$, the action density, and
$G(p_{\min})$ are reserved for validation rather than included in the fit.
We validate the kernel using held-out blocked-versus-direct distributions,
comparing means, widths, and two-sample Kolmogorov--Smirnov statistics, and
we test transfer from $L=32\to16$ to $L=64\to32$.
Fourier-space bounds on $K(p)$ and $K^{-1}(p)$ are imposed separately as
invertibility and conditioning checks.

We found several smoothing kernels with comparable performance.
The numerical results presented below use the $7\times7$ kernel listed in
Table~\ref{tab:phi4_kernel_orbits}.  A more compact $5\times5$ kernel also
gives similar matching quality, while increasing the spatial footprint beyond
$7\times7$ did not produce sufficient improvement to justify the additional
complexity.

 \begin{table}[tbh]
  \centering
  \caption{
    Optimized $7\times7$ $\phi^4$ blocking kernel, represented by
    $D_4$ symmetry orbits about the central site. The tabulated coefficients
    already include the anomalous-dimension normalization factor
    $2^{\eta/2}$.}
  \label{tab:phi4_kernel_orbits}
  \begin{ruledtabular}
  \begin{tabular}{ccc}
  Offset orbit $(|\Delta x|,|\Delta y|)$ & Multiplicity & $K(\Delta x,\Delta y)$ \\
  \hline
  $(0,0)$ & 1 & $ 0.888641822$ \\
  $(1,0)$ & 4 & $ 0.010508374$ \\
  $(1,1)$ & 4 & $-0.066254410$ \\
  $(2,0)$ & 4 & $ 0.037116968$ \\
  $(2,1)$ & 8 & $ 0.022657096$ \\
  $(2,2)$ & 4 & $-0.001649881$ \\
  $(3,0)$ & 4 & $ 0.020222068$ \\
  $(3,1)$ & 8 & $ 0.007299892$ \\
  $(3,2)$ & 8 & $-0.003848342$ \\
  $(3,3)$ & 4 & $-0.001693935$ \\
  \end{tabular}
  \end{ruledtabular}
  \end{table}

\begin{figure*}
      \centering
      \includegraphics[width=0.3\linewidth]{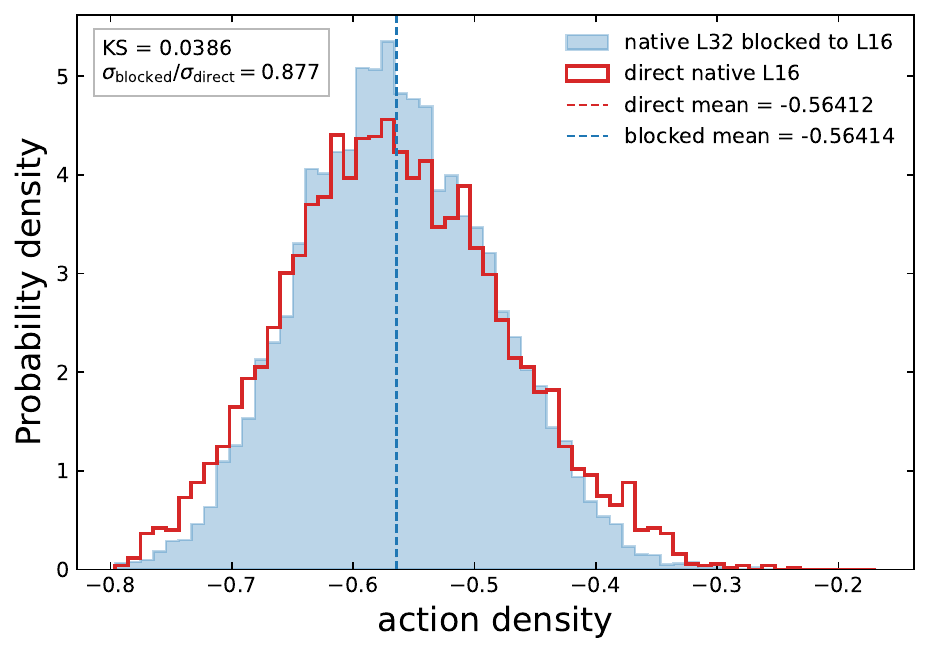} 
      \includegraphics[width=0.3\linewidth]{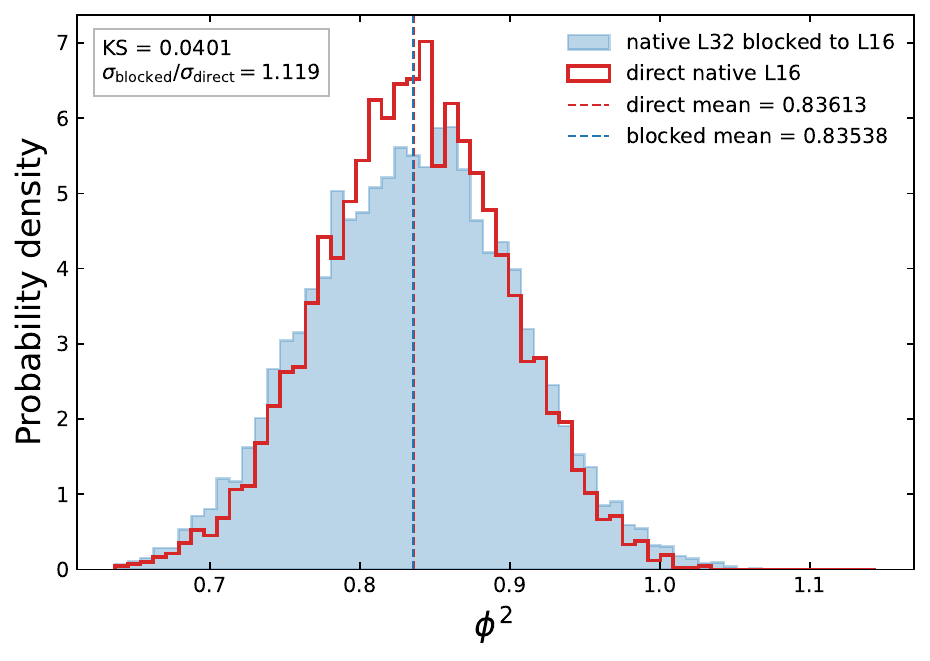} 
      \includegraphics[width=0.3\linewidth]{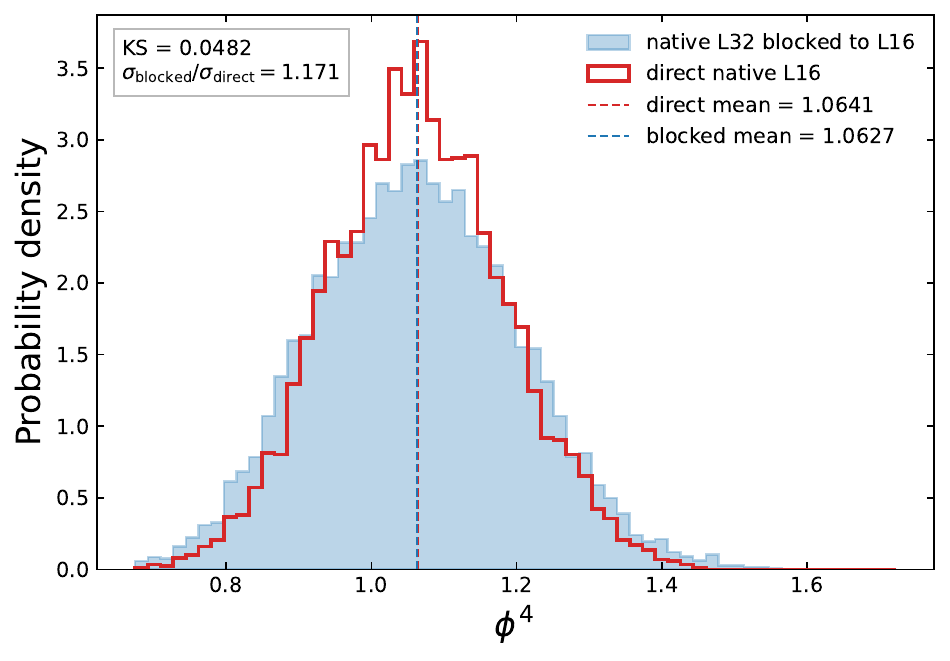} \\
      \includegraphics[width=0.3\linewidth]{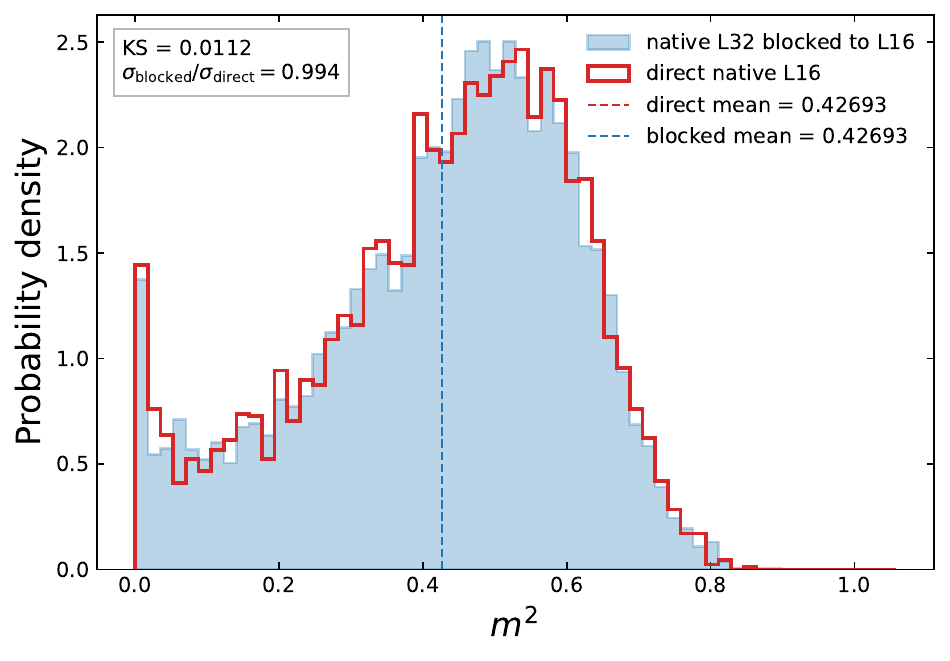}
      \includegraphics[width=0.3\linewidth]{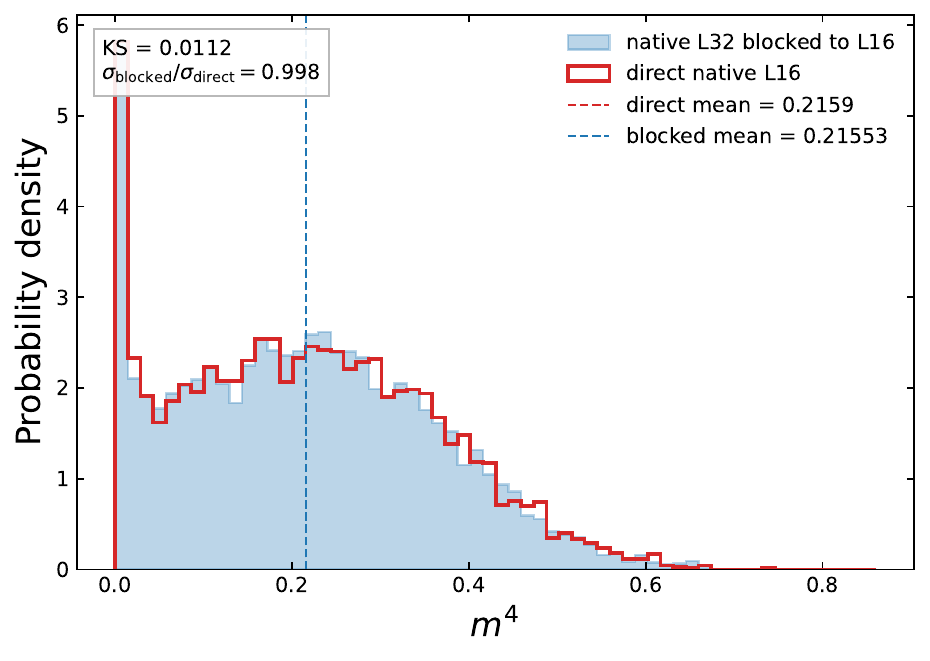}
      \includegraphics[width=0.3\linewidth]{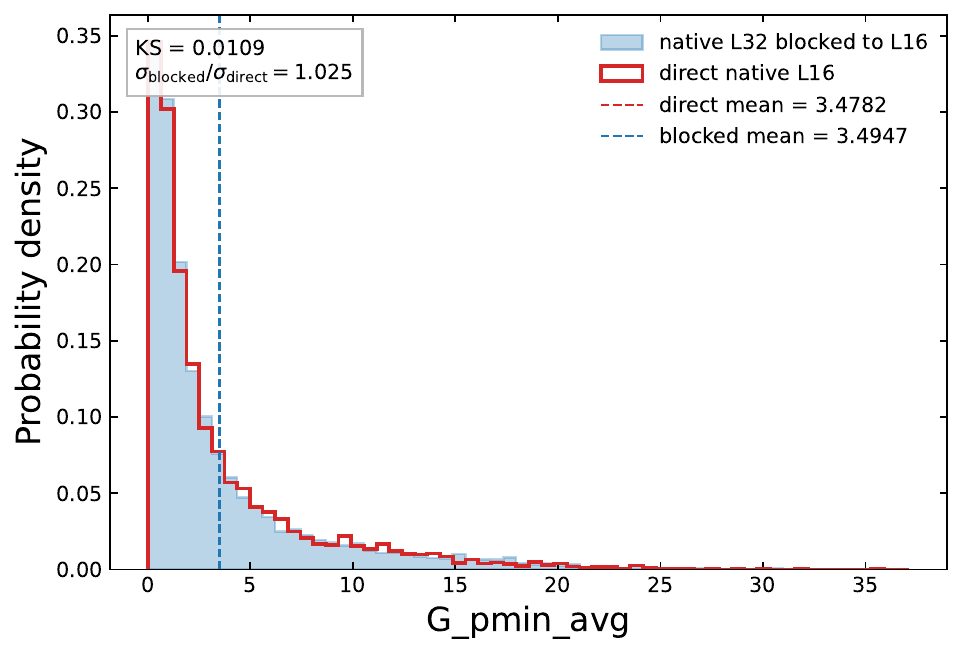}
      \caption{Approximate perfect blocking test for $L32\to L16$ blocking.  Direct coarse configurations and blocked fine configurations should have overlapping  distributions. From left to right, the panels compare the action densities, $\langle \phi^2\rangle$ and $\langle \phi^4\rangle$ distributions, and the long distance quantities   $m^2$ and $m^4$ and $G(p_\mathrm{min})$. }
      \label{fig:L32-to-16-blocking-histograms}
\end{figure*}

The conditioning factor of the kernel is defined in Fourier space
\begin{equation}
  \kappa_K=\frac{\max_{\mathbf p}|K(\mathbf p)|}
  {\min_{\mathbf p}|K(\mathbf p)|}.
\end{equation}
For the selected kernel,
\begin{equation}
  \min_{\mathbf p} K(\mathbf p)=0.54372,\qquad
  \max_{\mathbf p} K(\mathbf p)=1.24390,
\end{equation}
so
\begin{equation}
  \kappa_K=2.2878,
  \qquad
  \max_{\mathbf p}|K(\mathbf p)^{-1}|=1.8391.
\end{equation}
  The nonzero minimum is the key invertibility check; the modest condition number means inversion does not strongly amplify modes.

  \begin{table}[tb]
  \centering
  \caption{
  Perfect-blocking comparison on the $L=16$ coarse lattice using the
  $7\times7$ kernel of Table~\ref{tab:phi4_kernel_orbits}.
  Direct native $L=16$ configurations are compared with native $L=32$
  configurations blocked to $L=16$. The KS column gives the two-sample
  Kolmogorov--Smirnov statistic defined in \cref{app:statistics}; the final column is the ratio of blocked
  to direct standard deviations.
  }
  \label{tab:perfect_blocking_L16}
  \begin{ruledtabular}
  \begin{tabular}{lcccc}
  Operator & Direct $L=16$ & Blocked $32\to16$ & KS & Std. ratio \\
  \hline
  $S/V$ & $-0.5641(13)$ & $-0.56414(79)$ & $0.0386$ & $0.877$ \\
  $\phi^2$ & $0.83613(86)$ & $0.83538(68)$ & $0.0401$ & $1.119$ \\
  $\phi^4$ & $1.0641(17)$ & $1.0627(14)$ & $0.0482$ & $1.171$ \\
  $\langle\phi^4\rangle/\langle\phi^2\rangle^2$
   & $1.5220(10)$ & $1.52103(65)$ & $0.0194$ & $0.953$ \\
  $\mathrm{NN}$ & $1.1639(27)$ & $1.16295(19)$ & $0.0160$ & $1.028$ \\
  $2\mathrm{nn}$ & $0.9945(35)$ & $0.99413(25)$ & $0.0099$ & $0.995$ \\
  $\mathrm{diag}$ & $0.5267(16)$ & $0.52544(11)$ & $0.0165$ & $1.001$ \\
  $m^2$ & $0.4269(26)$ & $0.4269(18)$ & $0.0112$ & $0.994$ \\
  $m^4$ & $0.2159(20)$ & $0.2155(14)$ & $0.0112$ & $0.998$ \\
  $G(p_{\min})$ & $3.478(61)$ & $3.495(44)$ & $0.0109$ & $1.025$ \\
  \end{tabular}
  \end{ruledtabular}
  \end{table}

An essential diagnostic of the blocking transformation is the overlap between
blocked fine ensembles and directly generated (or ``native'' for short) coarse ensembles.
Figures~\ref{fig:L32-to-16-blocking-histograms} and
\ref{fig:L128-to-64-blocking-histograms} show representative distributional comparisons at
$32\to16$ and $128\to64$, while Tables~\ref{tab:perfect_blocking_L16}, \ref{tab:perfect_blocking_L32}, and \ref{tab:perfect_blocking_L64} quantify the
matching over all three successive volume pairs. They list expectation values, two-sample
Kolmogorov--Smirnov distances as defined  in \cref{app:statistics}, and the width ratio
\begin{equation}
R_\sigma=\frac{\sigma_{\rm blocked}}{\sigma_{\rm direct}} .
\end{equation}
Agreement is particularly strong for the long-distance observables
$m^2$, $m^4$, and $G(p_{\min})$, while the local distributions show somewhat
larger but still controlled differences.  The KS distances are typically at
the few-percent level and the width ratios remain close to unity.  Most
importantly, the agreement persists from $32\to16$ through $128\to64$,
although the kernel was optimized only on $32\to16$, providing a direct test
of the volume transfer required for recursive cascade upscaling.

\subsection{MCRG test of the blocking transformation} \label{sec:mcrg-test} 

As an independent test of the approximate perfect-blocking transformation, we apply the Monte Carlo renormalization-group (MCRG) method of Swendsen~\cite{Swendsen:1979gn}.  The method determines the linearized RG transformation matrix from correlations among operators measured before and after blocking, without requiring an explicit parameterization or simulation of the blocked action.

For a scale-$s$ transformation, the eigenvalues of the linearized RG transformation matrix at the fixed point are
\begin{equation}
    \lambda_i=s^{y_i},
\end{equation}
where $y_i$ are the RG scaling exponents.  In the even sector of the two-dimensional Ising critical point, there is a single relevant exponent, $y_t=1/\nu$, while the remaining even eigenoperators are irrelevant, with $y_i<0$.  Away from the fixed point, estimates obtained from successive blocking levels generally drift, with additional systematic effects from finite volume and truncation of the operator basis.  Details of the MCRG analysis are summarized in App.~\ref{app:mcrg}.

 We determine the leading even eigenvalue $\lambda_t$ using a five-operator basis containing the $\mathbb{Z}_2$-even local quadratic and quartic interactions, together with short-range bilinear operators.  The anomalous dimension $\eta=1/4$ is not determined in this analysis; it enters through the normalization $2^{\eta/2}$ of the blocked field used in constructing the approximate perfect kernel.

Table~\ref{tab:mcrg-nu} compares the resulting estimates of $\nu$ for the optimized $7\times7$ kernel with those obtained from a simple $2\times2$ block average using the same overall field normalization.  Each row starts from an independently generated native critical ensemble, and moving to the right corresponds to additional RG blocking steps.

 \begin{table}[t] 
 \centering 
 \caption{
Thermal exponent $\nu$ obtained from the leading even eigenvalue of the linearized RG transformation matrix~\cite{Swendsen:1979gn}.  The exact value for the two-dimensional Ising universality class is $\nu=1$.  Dashes denote blocking levels not available for the indicated starting volume.  The final $16\to8$ step is retained as a finite-volume diagnostic.  The rapid stabilization of the $7\times7$ results near $\nu=1$ contrasts with the slower drift observed for the simple block average.
}
 \label{tab:mcrg-nu} 
 \begin{tabular}{c|cccc} \hline\hline Starting $L$ & $128\to64$ & $64\to32$ & $32\to16$ & $16\to8$ \\ 
 \hline \multicolumn{5}{c}{ $2\times2$ averaging kernel} \\ $32$ & -- & -- & $0.7209(68)$ & $0.7893(52)$ \\ $64$ & -- & $0.7325(110)$ & $0.7856(72)$ & $0.8411(77)$ \\ $128$ & $0.7203(96)$ & $0.7817(63)$ & $0.8414(70)$ & $0.9253(126)$ \\ 
 \hline \multicolumn{5}{c}{$7\times7$ approximate perfect kernel} \\ $32$ & -- & -- & $0.9416(67)$ & $1.0014(97)$ \\ $64$ & -- & $0.9392(85)$ & $0.9760(107)$ & $1.0060(133)$ \\ $128$ & $0.9444(82)$ & $0.9867(90)$ & $0.9945(106)$ & $1.0121(139)$ \\ 
 \hline\hline 
 \end{tabular} 
 \end{table} 

 The difference between the two transformations is substantial.  For the simple $2\times2$ averaging, the estimated $\nu$ drifts strongly under successive blocking and approaches $\nu=1$ only slowly.  In contrast, the $7\times7$ transformation gives $\nu\simeq0.94$ already on the first blocking step, essentially independent of whether the starting lattice has $L=32$, 64, or 128, and after one additional blocking step the estimate is within $1$--$2\%$ of the exact Ising value.

Thus, within the operator basis used here, the leading thermal eigenvalue of the $7\times7$ transformation is already close to its fixed-point value before any prior blocking and rapidly stabilizes under further RG steps.  This indicates that the fixed point of the $7\times7$ transformation lies close to the nearest-neighbor critical action used in our simulations.  The simple block average, by comparison, requires several RG iterations to approach the same regime.  Because the blocking kernel was optimized solely through blocked-versus-direct distribution matching and not using this MCRG criterion, the rapid approach of the thermal eigenvalue to its universal value provides an independent validation of the construction.

\section{Upscaling}
\label{sec:upscaling}

Once the smoothing kernel has been fixed, the smoothed field
$\psi=K\phi$ is represented in coarse/detail coordinates,
\begin{equation}
  \psi=(\phi_b,d_{01},d_{10},d_{11}),
\end{equation}
with $\phi_b=\psi_{\ee}$ the retained even-even residue class and the
remaining three residue classes as the detail variables.  The corresponding
fine field is reconstructed by
\begin{equation}
  \phi=K^{-1}\psi .
  \label{eq:inverse-kernel}
\end{equation}

This representation separates the two possible sources of error in the
inverse construction.  A mismatch between the blocked distribution
$\phi_b$ and the directly generated coarse distribution $\phi_c$ reflects
imperfect coarse--fine matching, whereas a mismatch in the conditional
distribution of the detail fields at fixed coarse field reflects the
conditional-flow approximation.  The two effects can therefore be diagnosed
separately.

\subsection{Conditional upscaling flow}
\label{sec:conditional-flow}

\begin{figure*}
      \centering
      \includegraphics[width=0.3\linewidth]{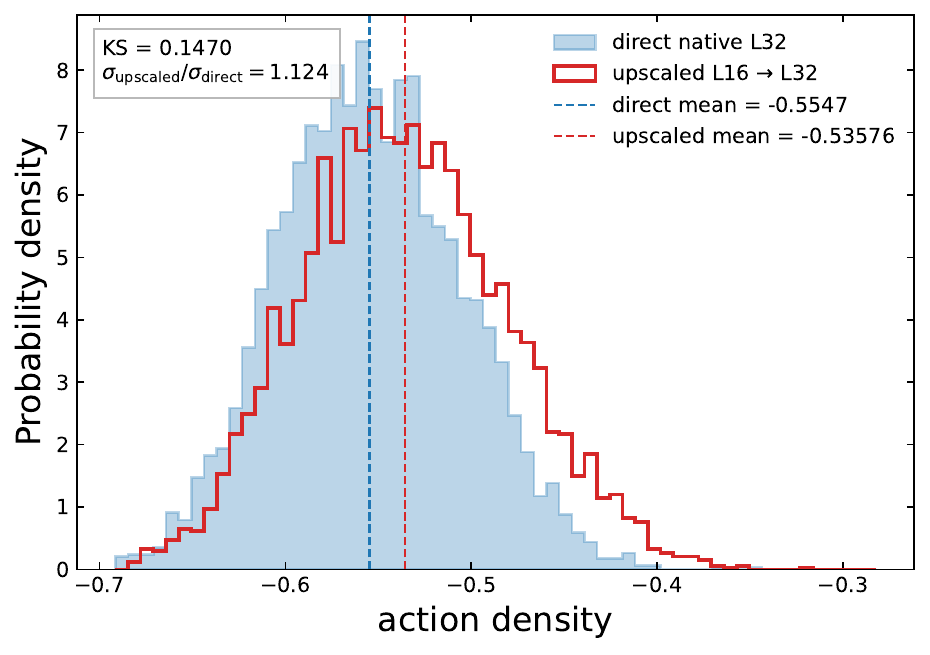} 
      \includegraphics[width=0.3\linewidth]{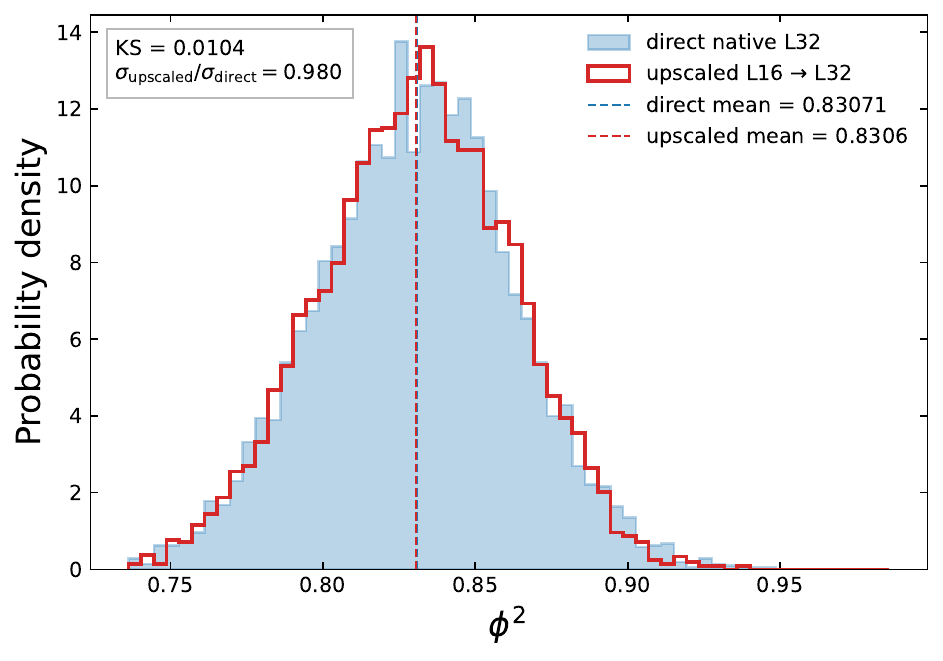} 
      \includegraphics[width=0.3\linewidth]{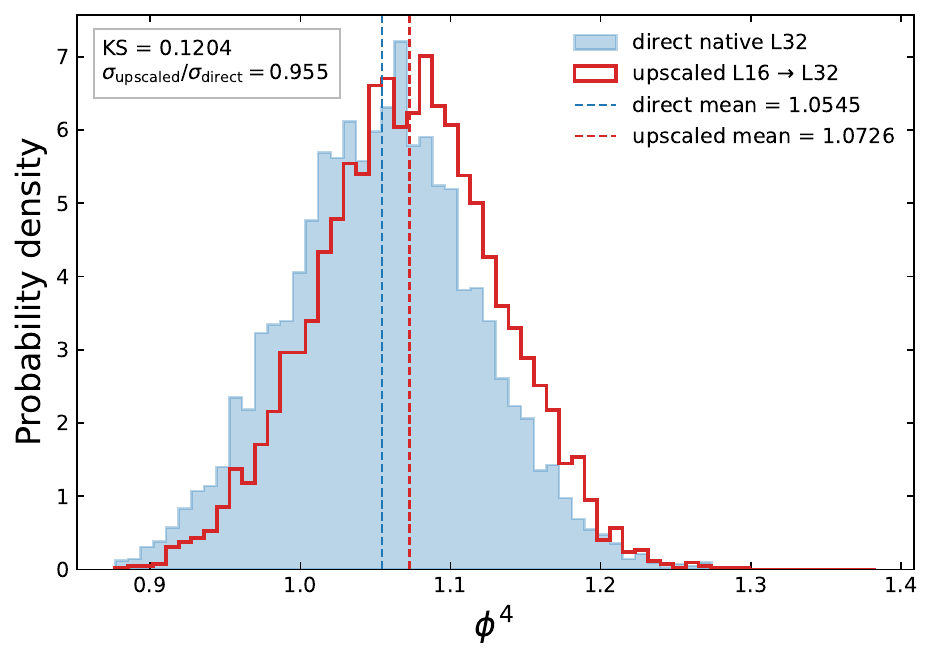} \\
      \includegraphics[width=0.3\linewidth]{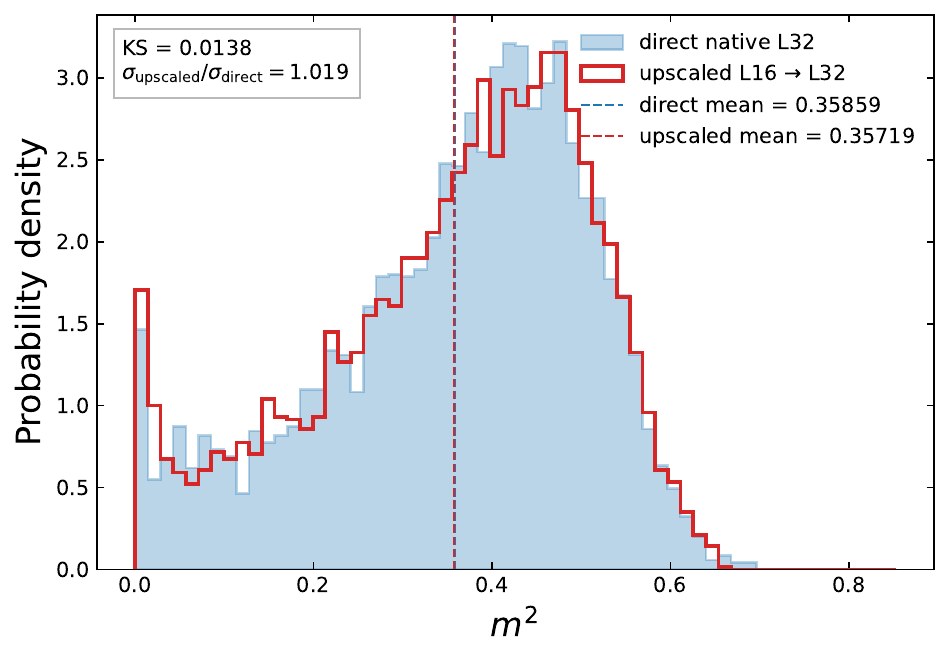}
      \includegraphics[width=0.3\linewidth]{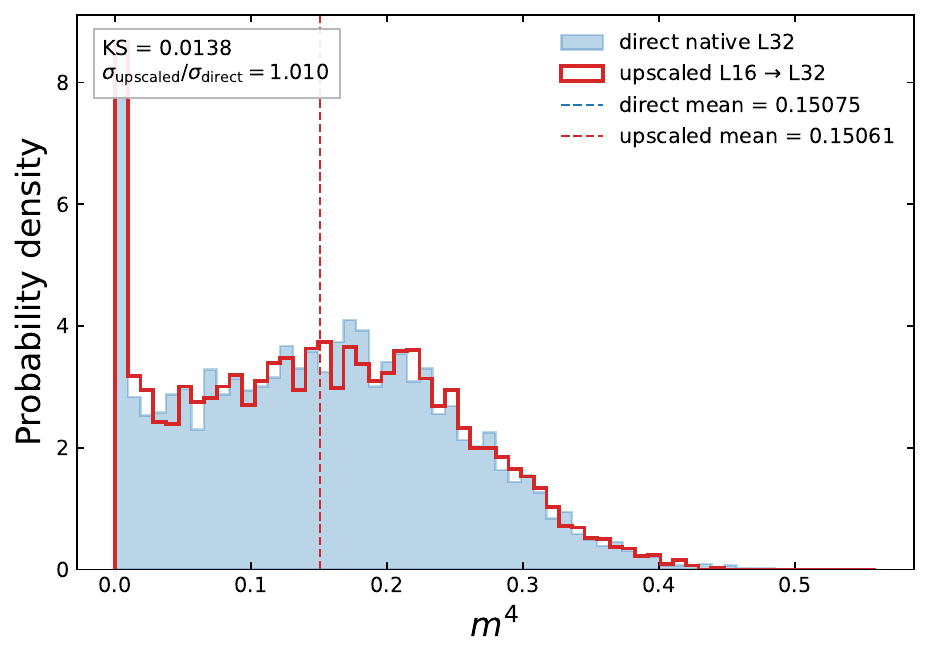}
       \includegraphics[width=0.31\linewidth]{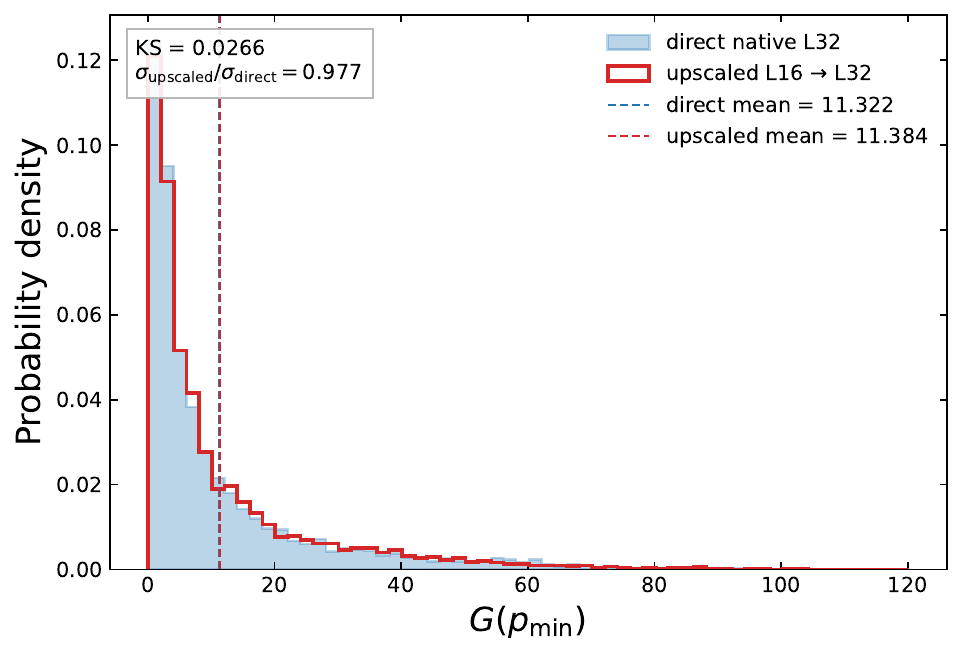}
      \caption{Initial conditional-flow proposal for $L=16\to32$.
The proposal already reproduces the long-distance distributions well, while
visible residual mismatch remains in local observables.  This mismatch is
removed efficiently by fine-action HMC rethermalization, but the raw
proposal is not accurate enough for an efficient global independence
Metropolis correction.  From left to right, the panels compare the action
density, $\langle\phi^2\rangle$, $\langle\phi^4\rangle$, and the long-distance
quantities $m^2$, $m^4$  and $G(p_\mathrm{min})$.
      }
      \label{fig:L16-to-32-initial-upscale}
\end{figure*}

The conditional upscaling proposal starts from a coarse configuration
\begin{equation}
  \phi_c\sim p_c(\phi_c)\propto e^{-S_c(\phi_c)} .
\end{equation}
Conditioned on $\phi_c$, the three detail sectors are then generated
sequentially with a conditional normalizing flow~\cite{Papamakarios:2021,Singha:2026slc},
\begin{align}
  d_{01} &\sim q_{\theta,01}(d_{01}\mid \phi_c),\\
  d_{10} &\sim q_{\theta,10}(d_{10}\mid \phi_c,d_{01}),\\
  d_{11} &\sim q_{\theta,11}(d_{11}\mid \phi_c,d_{01},d_{10}).
\end{align}
The resulting smoothed field is mapped back to the fine variables with
$K^{-1}$,
\begin{equation}
  \phi_{\rm prop}
  = K^{-1}F_\theta(\phi_c,z),
  \qquad z\sim p_0(z),
  \label{eq:upscale-flow}
\end{equation}
where $F_\theta$ denotes the conditional flow generating the complete
$\psi$ field while keeping its even-even component fixed to $\phi_c$.

For training, each fine configuration is transformed to coarse/detail
coordinates using the invertible smoothing map,
\begin{equation}
  \psi = K\phi_f,\;
  \phi_b=\psi_{ee},\;
  d=(d_{01}, d_{10}, d_{11}) = (\psi_{eo},\psi_{oe},\psi_{oo}),
  \label{eq:conditional-flow-decomposition}
\end{equation}
so that the training data consist of matched pairs $(\phi_b,d)$.
The conditional density is factorized over the three detail sectors,
\begin{align}
  q_\theta(d\mid\phi_b)
  =
  &q_{\theta,01}(d_{01}\mid\phi_b)\,
  \cdot q_{\theta,10}(d_{10}\mid\phi_b,d_{01})
  \notag\\
  &\cdot
  q_{\theta,11}(d_{11}\mid\phi_b,d_{01},d_{10}).
  \label{eq:conditional-flow-factorization}
\end{align}
The two edge sectors are generated first, followed by the body-centered
sector.  Each conditional factor is represented by rational-quadratic spline
(RQS) transformations with convolutional conditioners, preserving translation
equivariance while allowing each sector to depend on the coarse field and on
the previously generated detail sectors.  Alternatively, we have tried a second conditional flow setup with slight variations: the three sectors are trained simultaneously, and alternating affine and RQS layers are used, with $\mathbb{Z}_2$ equivariance imposed by making the coupling layers odd in $\phi$.  This second setup is used on the largest-volume upscaling shown below.

The first conditional model, $q_{\theta,01}$, is initialized independently.
The $q_{\theta,10}$ and $q_{\theta,11}$ models are initialized from the
best-validation-NLL checkpoint of the preceding stage, and each stage is then
optimized with a fresh optimizer.  The production flow is trained exclusively
by conditional maximum likelihood,
\begin{equation}
  \mathcal{L}_{\rm NLL}(\theta)
  =
  -\left\langle \log q_\theta(d\mid\phi_b)\right\rangle_{\rm train}.
  \label{eq:conditional-flow-nll}
\end{equation}
Model selection is based on validation NLL.  Observable distributions,
including local moments, action density, and kurtosis-sensitive quantities,
are used only as held-out diagnostics and in subsequent proposal and
rethermalization tests; they do not enter the training objective.  

Figure~\ref{fig:L16-to-32-initial-upscale} compares the raw conditional-flow
proposal with a native ensemble before any fine-action rethermalization.  The
main feature is a clear separation between local and infrared observables.
Local distributions, especially the action density and $\langle\phi^4\rangle$,
show visible shifts relative to the native target, while the long-distance
magnetization moments remain much better reproduced.

The same flow is applied without retraining at larger volumes.  Additional
raw-upscaling histograms for $L=32\to64$ and $L=64\to128$ are shown in
Appendix~\ref{app:extra-plots}, Figs.~\ref{fig:L32-to-64-initial-upscale}
and~\ref{fig:L64-to-128-initial-upscale}.  The local mismatch increases as
the flow is applied farther from its training volume.  Although the detail
fields represent predominantly short-distance degrees of freedom, the trained
conditional model is still a finite-volume approximation to their exact
conditional distribution.  In particular, the convolutional conditioners are
trained with periodic boundary conditions on the training lattice, so applying
the same network on a larger periodic volume need not reproduce the exact
larger-volume conditional distribution.
The resulting raw proposal is therefore not accurate enough for an efficient
global independence Metropolis correction, whose acceptance is sensitive to
extensive action differences.

By contrast, the infrared observables are much less affected.  The
magnetization moments $m^2$ and $m^4$ remain close to the native distributions
even when the local action-density distribution is significantly displaced.
This is the desired behavior: the matched coarse field supplies the infrared
structure, while the conditional flow fills in the missing ultraviolet detail
variables with sufficient accuracy so that the remaining mismatch can be removed
by a short fine-action rethermalization.

Retraining or adapting the conditional flow at each larger volume would likely
reduce the visible local discrepancies in the raw proposal.  We do not pursue
this here because the residual mismatch produced by the volume-transferred flow
is removed efficiently by the fine-action rethermalization described in
Sec.~\ref{sec:mh-updates}.  The conditional flow is therefore used as an
RG-guided initializer rather than as a standalone exact sampler.

\section{Rethermalization}
\label{sec:mh-updates}
\subsection{Rethermalization strategy}
\label{sec:action_correction}
\begin{figure*}[t]
\centering
\includegraphics[width=0.98\textwidth]{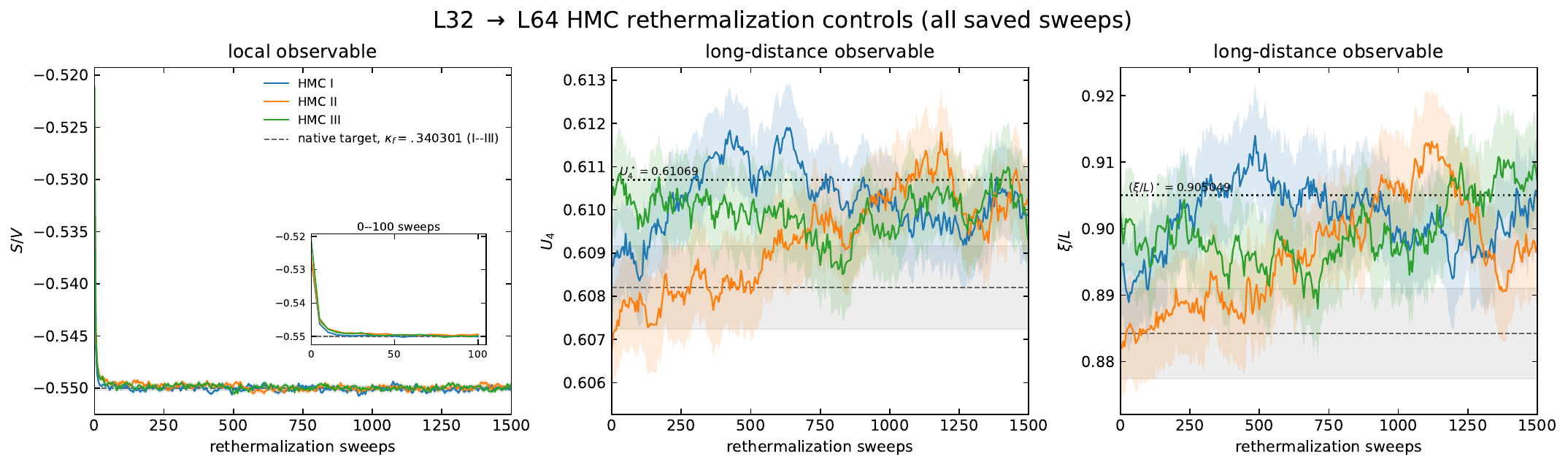}
\caption{ Controlled $L=32\to64$ HMC rethermalization tests to the same critical
fine action, $(\lambda_f,\kappa_f)=(1,0.340301)$.
HMC~I uses matched critical coarse and fine actions,
$(\lambda_c,\kappa_c)=(1,0.340301)$;
HMC~II starts from a thermally mistuned coarse ensemble,
$(\lambda_c,\kappa_c)=(1,0.340100)$;
HMC~III starts from a different microscopic critical action,
$(\lambda_c,\kappa_c)=(0.9,0.342410)$.
The left panel shows the local action density $S/V$, while the middle and
right panels show the long-distance observables $U_4$ and $\xi/L$.
All three cases rapidly approach the target value in the local observable,
whereas the long-distance observables show slow relaxation only for the
thermally mistuned HMC~II case.
}
\label{fig:hmc-rethermalization-controls}
\end{figure*}
The conditional upscaling flow described in
Sec.~\ref{sec:upscaling} generates fine-lattice configurations whose
long-distance modes are inherited from the matched coarse ensemble.  As seen
in the raw-proposal tests, the dominant residual discrepancies occur in local
quantities such as the action density, field moments, and short-distance
correlations, while infrared observables are already much more accurately
reproduced.  The final stage of the construction is therefore a fine-action
rethermalization that removes this residual mismatch.

In principle, the residual mismatch could instead be corrected by a global
Metropolis--Hastings accept/reject step, as in standard flow-based sampling
\cite{Albergo:2019eim}.  For an independence proposal, however, the acceptance
depends on the log probability ratio between the proposal and target
distributions.  Even small local errors accumulate over the lattice volume,
so fluctuations of this ratio grow with $V$ and global acceptance deteriorates
rapidly unless the proposal becomes increasingly accurate.  This is precisely
the unfavorable volume scaling that the present construction is intended to
avoid.

We therefore use the conditional flow as an initializer rather than as a
standalone exact sampler.  After upscaling, the proposed fine configuration
$\phi_{\rm prop}$ is evolved with a Markov update whose stationary
distribution is the target fine action $S_f$.  We study both Hybrid Monte
Carlo (HMC) and Wolff-cluster evolution \cite{Wolff:1989,Luscher:2003vf}.
Given sufficient evolution, either algorithm converges to the correct target
distribution; the practical question is how rapidly the different components
of the residual mismatch relax \footnote{Rethermalization following RG-matched prolongation has been studied in gauge and gauge--fermion systems in Refs.~\cite{Detmold:2016rnh,Detmold:2018zgk}. In those studies the prolongated configurations retained residual overlap with both short-distance and slowly relaxing long-distance modes, with the rethermalization rate sensitive to the quality of the coarse--fine matching.}.
To distinguish residual short-distance mismatch from mismatch along the
relevant thermal direction, we perform the three controlled
$L=32\to64$ tests shown in Fig.~\ref{fig:hmc-rethermalization-controls}.
HMC~I is the matched critical case used in the standard cascade.
HMC~II deliberately mistunes the coarse thermal coupling,
$\kappa_c=0.340100$, while evolving toward the critical fine target
$\kappa_f=0.340301$.
HMC~III instead changes the microscopic critical action, using
$(\lambda_c,\kappa_c)=(0.9,0.342410)$, still on the critical surface ~\cite{Bosetti:2015lsa}, while keeping the fine target at
$(\lambda_f,\kappa_f)=(1,0.340301)$.
The latter changes the short-distance action substantially while preserving
critical long-distance physics.

The contrast among the three cases is clear.  The action density in the left
panel relaxes rapidly in all three tests, including HMC~III where the initial
microscopic mismatch is large.  The long-distance observables behave
differently.  In the matched case HMC~I and in the distinct-critical-action
case HMC~III, $U_4$ and $\xi/L$ show no comparable long relaxation.
By contrast, the deliberately mistuned HMC~II ensemble approaches the target
only over a much longer HMC time scale.

These results support the RG interpretation of the rethermalization stage.
A large mismatch in the microscopic action does not necessarily entail slow relaxation,
provided that the inherited infrared distribution is correct. In contrast, a mismatch along
the relevant thermal direction modifies the IR distribution
and therefore couples to a slowly relaxing critical mode. Precise matching
is thus what allows the standard  cascade to achieve rethermalization within a short time scale.

We also performed the reverse thermal-mistuning test, in which a critical
coarse ensemble is evolved with the slightly off-critical fine action
$\kappa_f=0.340100$.  This case, denoted HMC~IV in
Tables~\ref{tab:therm_l64_common_local} Appendix~\ref{app:rethermalization} and
\ref{tab:therm_l64_common_long}, shows the same qualitative separation:
local observables relax rapidly, while $U_4$, $\xi/L$, and the susceptibility
evolve on a much longer time scale. Thus the slow relaxation associated with
thermal mistuning is observed in both directions.

As a complementary test, we also rethermalize selected ensembles with Wolff
cluster updates.  Tables~\ref{tab:therm_l64_common_local} and
\ref{tab:therm_l64_common_long} in Appendix~\ref{app:rethermalization} compare
Wolff and HMC evolution for both the matched critical case and a deliberately
mistuned thermal coupling.  Both algorithms restore local observables rapidly.
For the thermally mistuned ensemble, however, the long-distance observables
relax much more quickly under Wolff evolution than under HMC.  This shows that
the long HMC transient in the mistuned case reflects slow dynamics along the
critical thermal direction rather than a failure of the upscaled proposal to
have support in the target ensemble.

The same tables also include the distinct-critical-action test discussed above,
for which HMC rapidly repairs the large microscopic mismatch without inducing
a comparably slow drift in infrared observables.  Taken together, these
comparisons support the RG interpretation that residual short-distance and
irrelevant-direction errors are inexpensive to remove, while a mismatch along
the relevant thermal direction can be much more persistent for a local
fine-action evolution.

For the remaining tests in this section we focus on HMC-based
rethermalization.  HMC is the more generally applicable update scheme and is
closer to the algorithms required for eventual extensions to gauge theories
and fermionic systems, where cluster updates of the type used here are not
available.

\subsection{HMC implementation and recursive cascade}
\label{sect:cascade}

\begin{figure*}[t]
    \centering
    \includegraphics[width=0.95\textwidth]{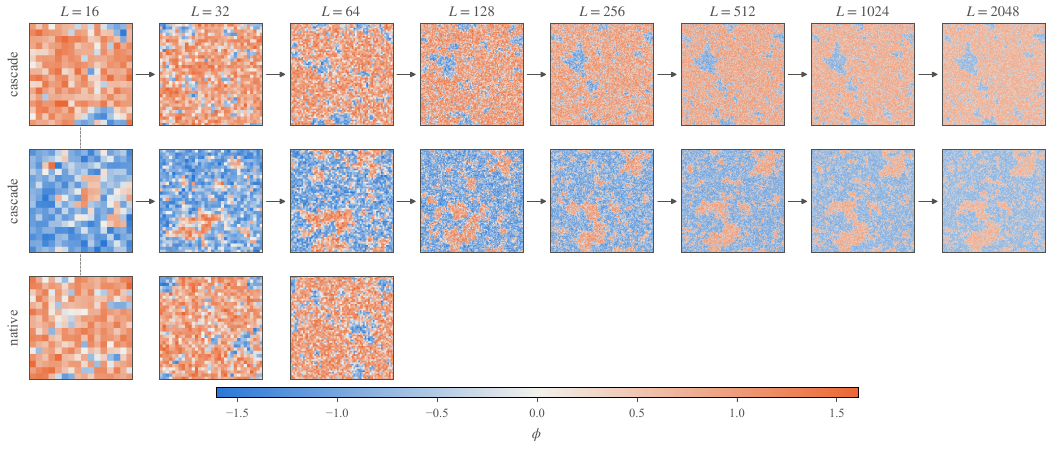} 
    \caption{Recursive inverse-blocking cascades in the two-dimensional $\phi^4$ model.
    Starting from equilibrated $L=16$ configurations at $\kappa_{\rm{cr}}$, each scale-two upscaling
    step is followed by HMC rethermalization targeting the fine action.
    Two independent cascades are shown, together with native small-volume
    configurations for visual reference.  The figure demonstrates the volume
    reach of the method; quantitative validation is provided by the
    rethermalization histories and expectation-value comparisons. 
    }
    \label{fig:upscale_tower}
\end{figure*}

We have tested two HMC implementations for fine-action rethermalization:
conventional full-volume HMC and domain-decomposed HMC (DD-HMC)
\cite{Luscher:2004pav}.  Both have the target fine action as their stationary
distribution and give broadly consistent rethermalization behavior, although we have found that DD-HMC may be less efficient unless the patch size and/or trajectory length are increased at larger volumes.  To be conservative and avoid possible contamination, we use full-volume HMC for all of the main results presented below.

For full-volume HMC, the parameters are tuned using standard procedures.
In the full-volume HMC runs we use trajectories of length $\tau=2$ and choose
the integration step size to maintain a Metropolis acceptance rate of
approximately $85\%$.  At fixed trajectory length, this requires decreasing
the step size with increasing volume, giving the familiar HMC cost scaling of
approximately $V^{5/4}$.  A trajectory length $\tau=2$ produces substantial
local evolution while maintaining a stable acceptance rate over the lattice
volumes studied.

For DD-HMC, the domains are arranged in a red--black checkerboard so that
nonoverlapping domains can be evolved in parallel.  Domain decomposition
provides a local and parallel implementation of the rethermalization step and
is particularly natural here because the residual mismatch produced by the
upscaling step is predominantly short-distance.  

The construction is applied recursively along the sequence
\begin{equation}
    L=16\to32\to64\to\cdots\to2048 .
\end{equation}
At every cascade level we use the same smoothing kernel, optimized from
$L=32\to16$ blocking, and the same conditional upscaling flow, trained on
the corresponding $L=16\to32$ inverse transformation.  The blocking kernel retains good matching quality on larger volumes, with particularly stable long-distance observables.
The raw conditional-flow proposal develops increasing short-distance mismatch
away from its training volume, but this remains sufficiently local that
retraining at each volume is unnecessary once rethermalization is included.

At each lattice level, the upscaled configuration is evolved for a fixed
number of HMC rethermalization sweeps before it is either used as the coarse
input to the next cascade step or retained for measurement.  The required
rethermalization length and its dependence on volume are tested explicitly
below.


The recursive construction introduces strong correlations between volumes
within a given cascade because all descendants inherit the infrared structure
of a common root configuration.  Measurements at different volumes in the same
cascade therefore cannot be treated as statistically independent, and
multivolume analyses resample entire cascades as described in
Appendix~\ref{app:statistics}.  At a fixed target volume, configurations
generated from independent root configurations remain independent.

This correlation is intrinsic to the construction: the expensive infrared
sampling is performed on the small root lattice, where decorrelation is cheap,
and the resulting long-distance structure is transported recursively to larger
volumes.  Consequently, extending a cascade to larger $L$ tests the stability
and volume transfer of the method but does not by itself provide additional
independent infrared statistics.

Figure~\ref{fig:upscale_tower} shows two representative cascades.
Starting from equilibrated $L=16$ configurations at $\kappa_{\rm cr}$, we
repeatedly apply scale-two upscaling followed by HMC rethermalization.
The construction remains stable over many blocking scales, reaching
$L=2048$ while preserving the long-distance structure inherited from the root
configuration.  Native small-volume configurations are shown only for visual
reference; quantitative validation is given below.

Figure~\ref{fig:cost} compares the computational scaling of direct HMC
generation with that of the cascade.  For direct critical HMC, the
autocorrelation time of long-distance observables grows approximately as
$\tau_{\rm int}\sim L^z$, with $z\simeq2.17$ for local dynamics at the
two-dimensional Ising critical point~\cite{nightingale1996dynamic}.  The cost
of generating a new statistically independent infrared configuration therefore
grows much faster than the lattice volume alone.

In the cascade, the infrared configuration is sampled at the root volume and
transported to larger lattices; the additional work at each level consists of
conditional upscaling and finite rethermalization.  As a rough measure of rethermalization time, we compute $\tau^\star(O)$ as the first-crossing time for a given observable $O$, which is the number of sweeps required for the central value of $O$ to fall within 1$\sigma$ of the final measurement after the full set of rethermalization sweeps.  The overall $\tau^\star$ shown in the plot is the largest (worst-case) value over the four observables $\{S/V, \langle \phi^2 \rangle, \langle \phi^4 \rangle / \langle \phi^2 \rangle^2, \langle nn \rangle \}$.  The points and error bars shown are a bootstrap median and 68\% confidence interval, showing a relatively flat volume dependence and that the choice of 500 sweeps is conservative at all volumes compared to the measured $\tau^\star$. 

Since the required number
of rethermalization sweeps varies only weakly with $L$, the marginal cost of
extending an existing cascade is dominated by the volume scaling of the local
computation.  On the laptop-scale resources used here, extending a cascade to
$L=2048$ requires only a few minutes, whereas extrapolating direct HMC
decorrelation to the same volume gives a time scale of order $10^2$ days per decorrelated configuration.

\begin{figure}[t]
\centering
\includegraphics[width=0.485\textwidth]{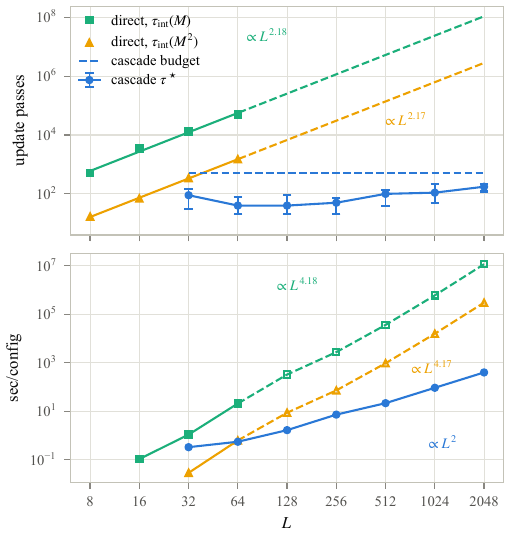}
\caption{Cost scaling of cascade upscaling vs. HMC. Solid symbols use measured autocorrelation times and generation costs; open symbols for HMC use measured generation costs per configuration, extrapolated using the autocorrelation times measured on smaller volumes.  In the top panel, $\tau^\star$ shows the estimated rethermalization time required for the cascade, as described in the text.  Our production runs shown here conservatively set 500 sweeps at all volumes, with correspondingly higher cost per configuration.}
\label{fig:cost}
\end{figure}

\subsection{Rethermalization tests}
\label{sec:rethermalization}
\begin{figure}[t]
\centering
\includegraphics[width=0.485\textwidth]{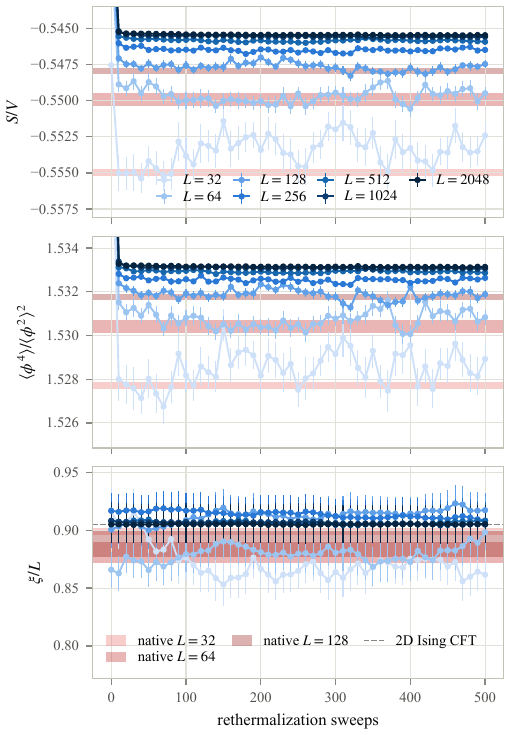}
\caption{
Rethermalization histories after conditional-flow initialization at several
cascade volumes.  We show the local action density $S/V$ and kurtosis $\langle \phi^4 \rangle / \langle \phi^2 \rangle^2$, as well as the
long-distance quantity $\xi/L$.  Where native reference ensembles are
available, their expectation values are indicated; at larger volumes the
late-sweep behavior provides a stationarity reference.  Visible transients
disappear well before the production choice of $n^\star=500$ HMC sweeps.
}
\label{fig:retherm-history}
\end{figure}

\begin{figure*}
      \centering
      \includegraphics[width=0.3\linewidth]{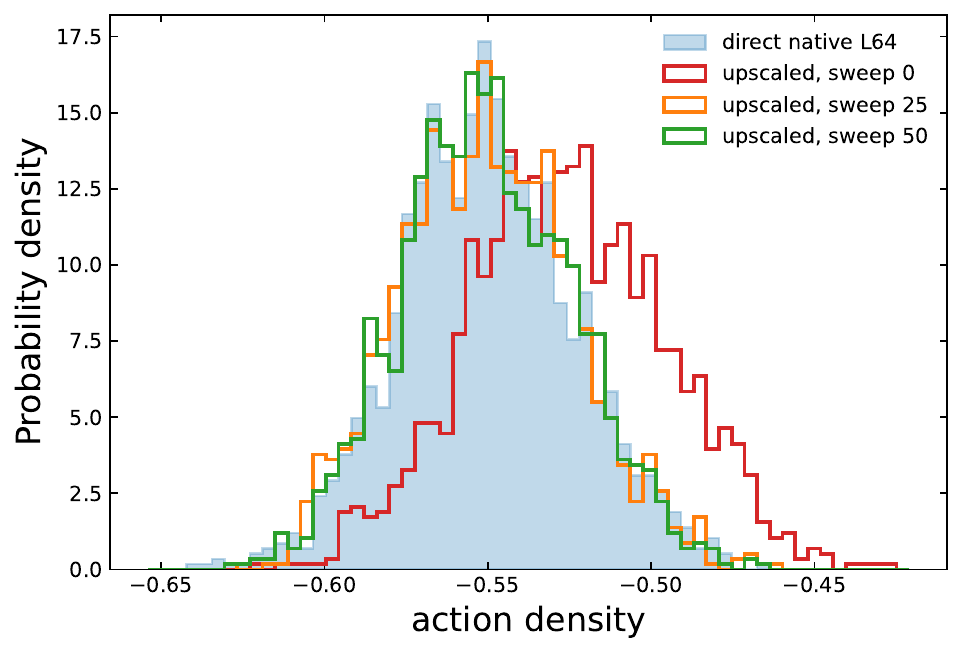} 
      \includegraphics[width=0.3\linewidth]{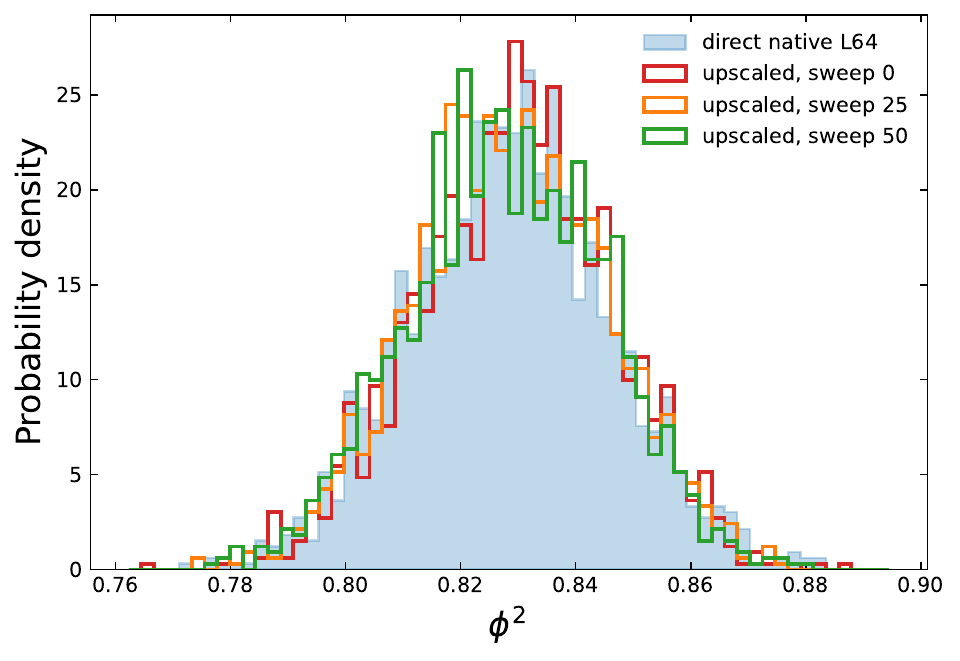} 
      \includegraphics[width=0.3\linewidth]{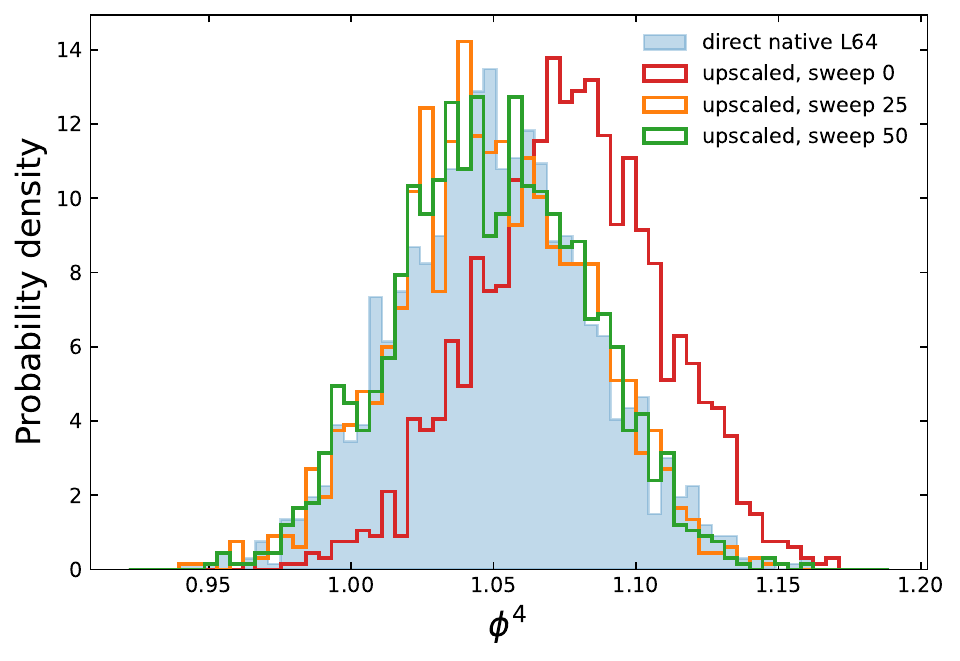} \\
      \includegraphics[width=0.3\linewidth]{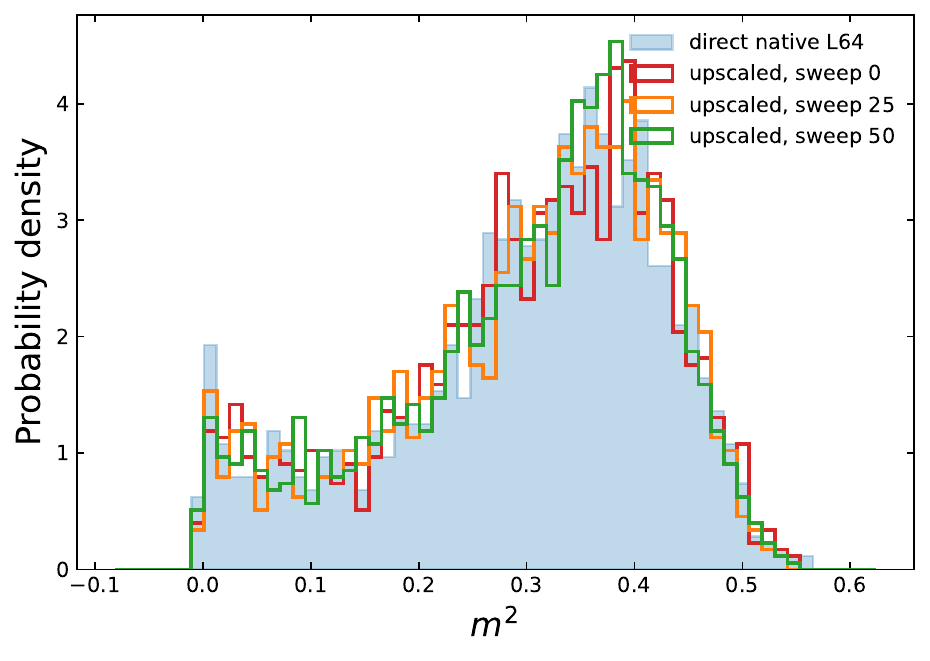}
      \includegraphics[width=0.3\linewidth]{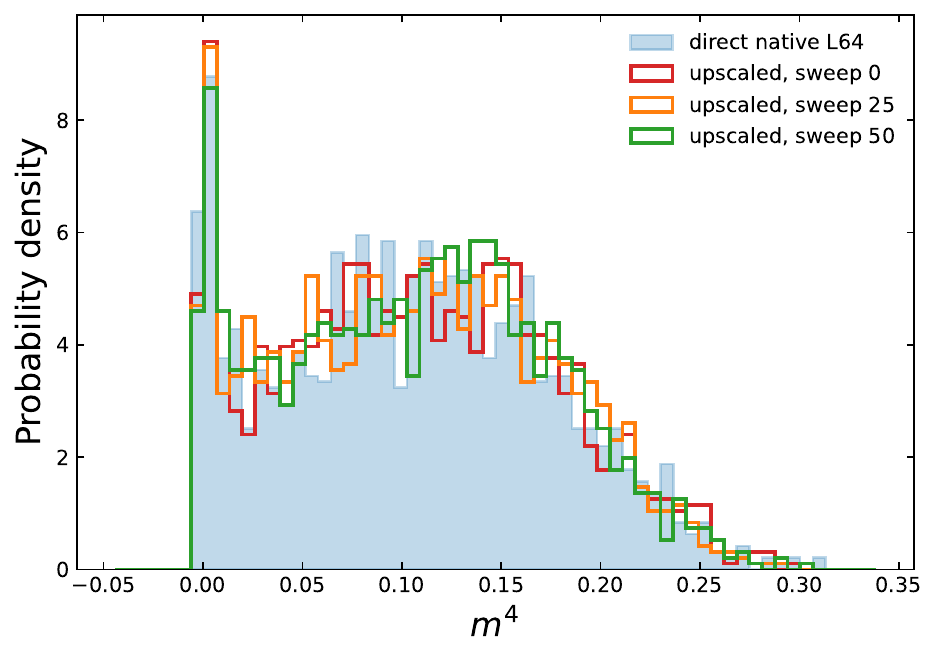}
      \includegraphics[width=0.3\linewidth]{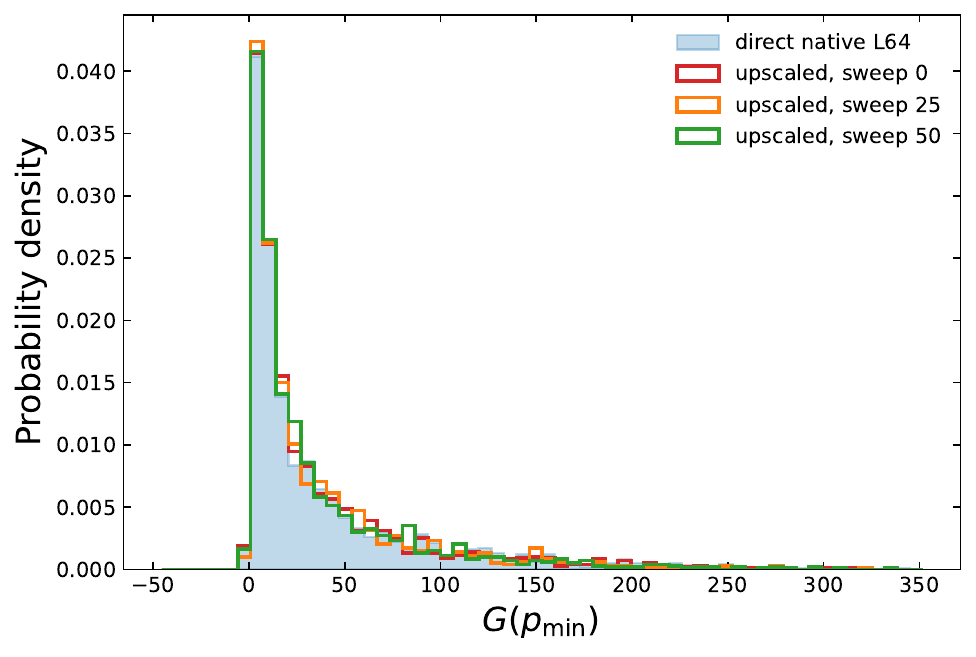}
      \caption{
Distributional rethermalization test for the $L=32\to64$ upscaling step.
The sweep-zero proposal shows the local mismatch discussed in
Fig.~\ref{fig:L32-to-64-initial-upscale}; the same ensemble is shown after
25 and 50 HMC sweeps.  Local distributions rapidly approach the native
reference, while the long-distance $m^2$ and $m^4$ distributions remain
stable.
}
      \label{fig:L32-to-64-rethermalization}
\end{figure*}

The purpose of these tests is to determine how much fine-action evolution is
required after each upscaling step and, whether that rethermalization time grows
with lattice volume.  This is not thermalization from a generic hot or cold
start: the upscaled configurations already inherit their long-distance
structure from the matched coarse ensemble, while the dominant residual
mismatch is short-distance.

We monitor both local and long-distance observables as functions of the number
of HMC sweeps.  Local quantities such as the action density and field moments
are particularly sensitive to imperfections in the detail-field proposal,
whereas $U_4$, $\xi/L$, $G(p_{\min})$, and the susceptibility test whether
the inherited infrared distribution remains stable during rethermalization.

Figure~\ref{fig:retherm-history} shows representative rethermalization
histories over several cascade volumes.  Where native ensembles are available,
the histories are compared directly with native expectation values.  At larger
volumes, where independent native ensembles are unavailable, the late-sweep
values provide a practical stationarity reference.  In all cases the visible
transients disappear well before the production rethermalization length.

Figure~\ref{fig:L32-to-64-rethermalization} illustrates the same relaxation
distribution for the $L=32\to64$ step.  The substantial sweep-zero mismatch
in local observables is largely removed by 25--50 sweeps, while the
long-distance distributions remain stable.  A corresponding
$L=64\to128$ comparison is shown in
Fig.~\ref{fig:L64-to-128-rethermalization} of
Appendix~\ref{app:extra-plots}.

Table~\ref{tab:retherm_L128} gives a quantitative summary for the
$L=64\to128$ step at sweeps 0, 25, and 50.  It complements the histograms by
showing that the large initial discrepancies in $S/V$ and $\phi^4$ are removed
rapidly, while the long-distance quantities remain statistically compatible
with the native ensemble throughout the rethermalization interval.

Across the volumes studied, the dominant local transient is largely removed
within approximately 20--50 HMC sweeps, and we resolve no significant growth
of this relaxation scale with $L$.  For production we nevertheless use the
conservative choice $n^\star = 500$
HMC sweeps after each upscaling step.  This choice lies well beyond the
observed transient region and provides substantial margin against residual
finite-rethermalization effects.

  \begin{table*}[t]
  \centering
  \caption{
  Rethermalization of the  L64$\to$L128 upscaled ensemble.
  Errors are bootstrap standard errors. The susceptibility is connected,
  $\chi=V(\langle m^2\rangle-\langle m\rangle^2)$, and $\xi/L$ is the
  second-moment correlation length.
  }
  \label{tab:retherm_L128}
  \begin{ruledtabular}
  \begin{tabular}{lcccc}
  Observable & Native & Sweep 0 & Sweep 25 & Sweep 50 \\
  \hline
  $S/V$
   & $-0.54791(20)$ & $-0.52474(42)$ & $-0.54751(36)$ & $-0.54791(36)$ \\
  $\phi^2$
   & $0.82699(13)$ & $0.82912(22)$ & $0.82679(22)$ & $0.82709(24)$ \\
  $\phi^4$
   & $1.04760(25)$ & $1.07655(42)$ & $1.04715(43)$ & $1.04793(46)$ \\
  $G(p_{\min})$
   & $128.5(2.4)$ & $126.1(4.3)$ & $124.7(4.3)$ & $125.5(4.3)$ \\
  $\chi$
   & $4127(25)$ & $4207(44)$ & $4159(44)$ & $4167(43)$ \\
   $U_4$
   & $0.6082(15)$ & $0.6113(27)$ & $0.6111(27)$ & $0.6111(27)$ \\
  $\xi/L$
   & $0.888(11)$ & $0.906(20)$ & $0.905(20)$ & $0.903(20)$ \\
  \end{tabular}
  \end{ruledtabular}
  \end{table*}

\section{Physics Results}
\label{sec:results}

As a final validation of the cascaded upscaling algorithm, we extract physical
quantities associated with the two-dimensional Ising critical point from the
generated ensembles.  Many exact or high-precision results are known for this
universality class, providing stringent benchmarks for the long-distance
physics.  Definitions of the observables used below are collected in
\cref{app:operators}.

\begin{figure}[t]
\centering
\includegraphics[width=0.485\textwidth]{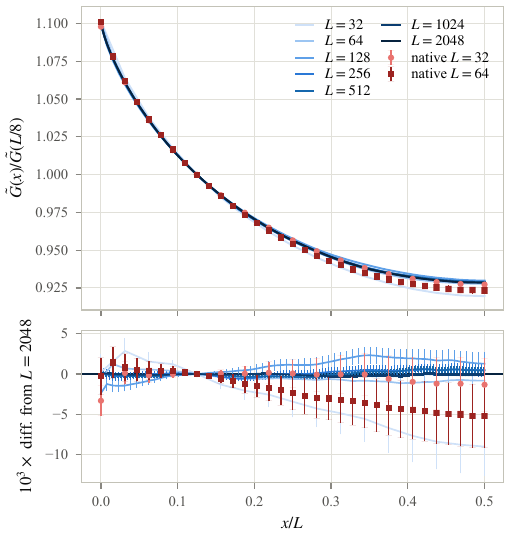}
\caption{Finite-size scaling collapse of the two-point correlation function $G(x)$
along the cascade.  Each curve is normalized to its value at $x=L/8$ and
plotted as a function of $x/L$.  Apart from small finite-size effects on the
smallest lattices, the cascade ensembles follow a common scaling function;
the native $L=64$ result is shown for comparison.}
\label{fig:curve_collapse}
\end{figure}

A simple qualitative test of finite-size scaling is curve collapse.  At a
scale-invariant point, two-point correlation functions measured on different
volumes become universal functions of the dimensionless separation $x/L$, up
to an overall normalization and subleading corrections to scaling.  In
Fig.~\ref{fig:curve_collapse} we show the zero spatial-momentum two-point function $G(t)$,
normalized to its value at $t=L/8$ on each ensemble.  Apart from small
finite-size deviations on the smallest lattices, the cascade results collapse
onto a common curve.  The native $L=64$ ensemble is consistent with the same
scaling function within the precision shown.

\begin{figure}[t]
\centering
\includegraphics[width=0.485\textwidth]{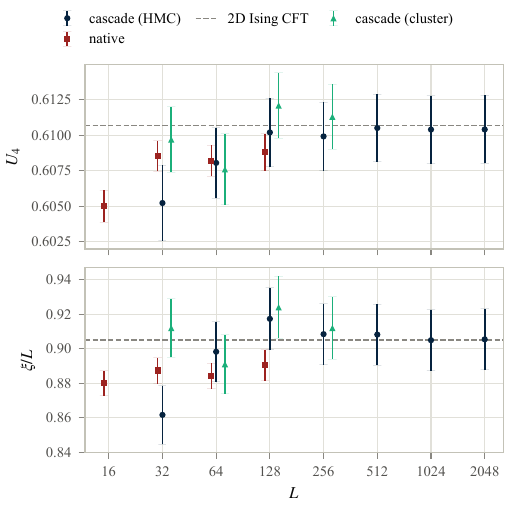}
\caption{Binder cumulant $U_4$ and normalized correlation length $\xi/L$ measured on
the cascade and native HMC ($L=16, 32$) and cascade ($L=64, 128$) ensembles.  The exact two-dimensional Ising
critical values from Ref.~\cite{Salas:1999qh} are shown for comparison.
Cascade points at different volumes are strongly correlated because they
inherit common infrared structure from the root ensembles.  Extended rethermalization results using the cluster algorithm (green triangles) are consistent with the HMC cascade.  Cascade results are shown for 2,000 independent chains; native results are for 10,000 configurations.}
\label{fig:invariants}
\end{figure}

We next consider two long-distance RG-invariant observables, the Binder
cumulant $U_4$ and the normalized correlation length $\xi/L$.  Their
infinite-volume critical values are known with high precision
\cite{Salas:1999qh},
\begin{equation}
    U_4^\star = 0.6106924(16),
    \qquad
    (\xi/L)^\star = 0.9050488292(4).
\end{equation}
Figure~\ref{fig:invariants} compares these values with measurements on the
cascade ensembles and on native HMC ensembles ($L=16, 32$) and native cluster ensembles ($L=64, 128$) at the smaller volumes.  Both
observables remain stable across the cascade and agree with the expected Ising
critical values within uncertainties.  
Because the different cascade volumes inherit the same long-distance
structure from common root configurations, these points are strongly
correlated.  Both observables  are consistent with the CFT predictions, showing a deviation of approximately $1\sigma$.

Critical exponents provide a more demanding test of finite-size scaling.  At
criticality, neglecting subleading corrections,
\begin{align}
    \chi &\propto L^{\gamma/\nu},\\
    \langle |m| \rangle &\propto L^{-\beta/\nu},\\
    \frac{d}{d\kappa}\log\langle m^2\rangle
        &\propto L^{1/\nu}.
\end{align}
We use these relations to extract $\gamma/\nu$, $\beta/\nu$, and $\nu$ from
the cascade ensembles, accounting for correlations among volumes in the fits
\cite{ferrenberg1991critical,Ferrenberg:2018zst}.

\begin{figure}[t]
\centering
\includegraphics[width=0.485\textwidth]{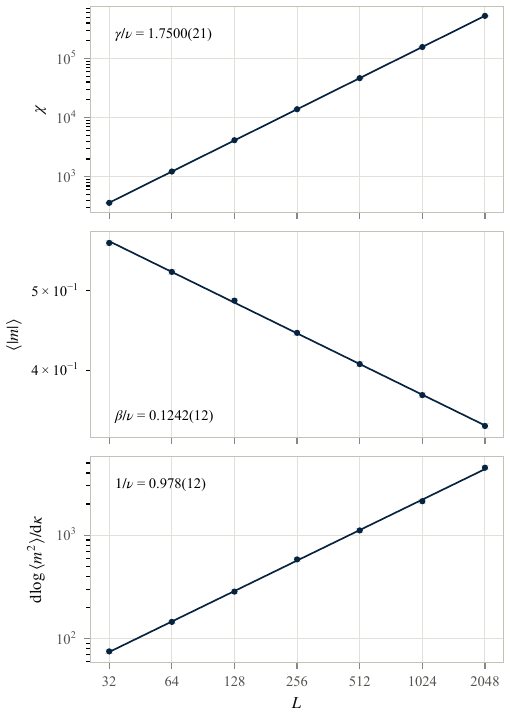}
\caption{Critical exponents extracted from finite-size scaling of the cascade
ensembles.  Fits account for correlations among volumes within the cascade.  Our determinations are all within $\sim 2\sigma$ of the exact exponents.}\label{fig:exponents}
\end{figure}

The resulting exponent estimates are shown in Fig.~\ref{fig:exponents}.  The
fits reproduce the expected Ising scaling behavior overall, with
$\nu$ consistent with the exact value $\nu=1$.  The other exponents $\gamma/\nu$ and $\beta/\nu$ are reproduced reasonably close to their exact values of $7/4$ and $1/8$ respectively.

We regard the exponent analysis as a consistency check on the finite-size scaling of the cascade. It is
not  competitive with other published determinations that use much higher statistics than our exploratory runs.

\section{Conclusions}
\label{sec:conclusions}

We have developed a renormalization-guided upscaling algorithm for lattice
field generation that separates the difficult infrared sampling problem from
the reconstruction of short-distance degrees of freedom.  An approximate
perfect-blocking transformation is optimized so that the RG-blocked fine
distribution is well represented by a simple coarse action.  Configurations
drawn from this matched coarse ensemble therefore supply the long-distance
structure, while a conditional normalizing flow reconstructs the detail fields
removed by blocking.  The resulting fine configuration is then evolved for a
short time with an update algorithm targeting the fine action in order to
remove the residual mismatch.

The two-dimensional $\phi^4$ theory at the Wilson-Fisher critical point provides a
stringent test of this construction.  The optimized blocking transformation
gives good agreement between directly generated coarse ensembles and blocked
fine ensembles over several successive volume pairs, and an independent MCRG
analysis shows that its leading thermal eigenvalue approaches the Wilson-Fisher
fixed-point value rapidly under blocking.  The conditional flow, trained only
on the $L=16\to32$ inverse transformation, can be reused recursively on much
larger lattices.  Although the raw proposal develops increasing
short-distance mismatch away from its training volume, the inherited
long-distance observables remain much more stable.

The rethermalization studies clarify why this construction is effective.
Residual local mismatch is removed rapidly, even when the microscopic source
action differs substantially from the target action, provided that the
inherited infrared distribution is correct.  By contrast, deliberately
mistuning the relevant thermal coupling produces much slower relaxation of
long-distance observables under HMC.  This separation provides direct evidence
that the cascade succeeds by transporting already-equilibrated infrared modes
from the coarse ensemble, leaving the fine-action evolution primarily to repair
short-distance structure.

Using the same blocking kernel and conditional flow throughout the cascade, we
have generated ensembles recursively from $L=16$ to $L=2048$.  The very large
volumes should be viewed primarily as a demonstration of the stability and
scale transfer of the construction: volumes within a given cascade remain
strongly correlated because they inherit the same infrared structure from a
common root configuration.  The computational advantage is therefore not that
each cascade level supplies new independent infrared information, but that
infrared configurations generated inexpensively on small lattices can be
transported to much larger volumes at modest additional cost.

The present study uses a scalar model for which efficient cluster algorithms
are available, but the rethermalization stage was deliberately implemented
with HMC and domain-decomposed HMC because these methods have much closer
analogs in gauge and fermion simulations.  The construction itself requires
three ingredients - a suitable blocking transformation, a conditional model for
the discarded degrees of freedom, and a fine-action evolution step - and none
is specific to scalar fields.  Extending these ingredients to compact spin
systems and gauge theories is therefore a natural next step toward applying
renormalization-guided upscaling in settings where critical slowing down is a
more serious computational obstacle.

\begin{acknowledgments}
We thank Dan Hackett, Jake Sitison, and Roman Marcarelli for useful discussions.  This research was partially supported by DOE grant DE-SC0010005.  Code development was performed with substantial assistance from OpenAI's ChatGPT and Codex and Anthropic's Claude Opus 5 and Fable 5.  A.H. and E.T.N. directed the development and carried out the debugging, validation, and testing of the final implementation.  A.H. acknowledges the stimulating environment at ECT* Trento during the June 2023 workshop ``Machine Learning for Lattice Field Theory and Beyond", which  inspired her work on inverse RG and normalizing flows.
\end{acknowledgments}

\appendix

\counterwithin{figure}{section}
\counterwithin{table}{section}

\renewcommand{\thefigure}{\thesection\arabic{figure}}
\renewcommand{\thetable}{\thesection\arabic{table}}

\section{Operators and estimators}
\label{app:operators}

This appendix compiles the definitions of the operators measured in this
study, together with the estimators employed for each and the corresponding
statistical analysis applied to them.  These operators and measurements are used throughout the paper, in particular for fitting and validating the blocking kernel, diagnosing the upscaling flow, monitoring rethermalization, and extracting physics results from the cascade-upscaled ensembles. 

\subsection{Conventions}

We work with $d=2$-dimensional square lattices with linear extent $L$ and periodic boundary conditions; each lattice has $V = L^d$ sites, with a single real scalar field $\phi_x$ at each site.  The action is given in \cref{eq:phi4-action}. 

Unless stated otherwise, quantities defined in terms of lattice fields are defined on a single field configuration.  Operator expectation values over an ensemble are denoted with angled brackets $\langle ... \rangle$.  

\subsection{Local operators}
\label{app:ops-local}

The ultralocal moments are the site-averaged even powers of the field,
\begin{equation}
  \phi^{2n} \equiv \frac{1}{V}\sum_{x} \phi_{x}^{2n} ,
  \qquad n = 1,2,3 ,
  \label{eq:app-moments}
\end{equation}
 The bilinear operators
at fixed separation $r$ are
\begin{equation}
  G(r) \equiv \frac{1}{\mathcal{N}_{r}} \sum_{r' \in \mathcal{O}_{r}}
              \frac{1}{V} \sum_{x} \phi_{x}\, \phi_{x+r'} ,
  \label{eq:app-twopoint}
\end{equation}
where $\mathcal{N}_{r}$ is the symmetry factor of the operator with respect to the rotational $D_4$ symmetry-group orbit $\mathcal{O}_r$. With periodic boundary conditions, the site-averaged correlator is identical for $r$ and $-r$ configuration by configuration, so only one representative of each pair is included in the orbit average.  In two dimensions, specific instances of $G(r)$ can be written instead as $G_{ij}$ where $i \geq j$ are the components of the vector $r$ defining the orbit; for example, $G_{10}$ is the nearest-neighbor bilinear which appears in the action.
The three shortest separations carry the names used in the tables,
\begin{equation}
  \mathrm{NN} = d\,\times G_{10} , \qquad
  \mathrm{2NN} = d\,\times G_{20} , \qquad
  \mathrm{diag} = G_{11} ,
  \label{eq:app-nn}
\end{equation}
where the factor of $d$ means that $\mathrm{NN}$ is the operator multiplying $-2\kappa$ in Eq.~\eqref{eq:phi4-action}.

The action density is defined as
\begin{equation}
  S/V \equiv \frac{1}{V}\, S[\phi] - \lambda ,
  \label{eq:app-actiondensity}
\end{equation}
where $S[\phi]$ is evaluated using the action of Eq.~\eqref{eq:phi4-action} at the value of the coupling at which the ensemble was generated. The
subtracted $\lambda$ corresponds to the constant term in $\lambda(\phi^{2}-1)^{2}$, which is
independent of the field and therefore omitted from the reported values; it induces a uniform shift in all $S/V$ entries
and hence cancels in any comparison.

\subsection{Magnetization sector}
\label{app:ops-magnetization}

The order parameter is the volume-averaged field,
\begin{equation}
  m \equiv \frac{1}{V}\sum_{x}\phi_{x} ,
  \label{eq:app-mag}
\end{equation}
and the moments reported are $m^{2}$, $m^{4}$ and $|m|$. No ensemble mean is subtracted; all moments are defined directly from
Eq.~\eqref{eq:app-mag}.

\subsection{The 2-point correlation function}
\label{app:ops-momentum}

The large-distance properties of the two-point function defined in
Eq.~\eqref{eq:app-twopoint} characterize the infrared dynamics of the system.
We use the zero-momentum projected correlation function
\begin{equation}
  \widetilde G(t) = \sum_{x_1} G(t,x_1),
  \label{eq:GT_zeromom}
\end{equation}
as well as the momentum-space correlation function
\begin{equation}
  G(p) = \frac{1}{V}\left| \sum_{x} e^{i p \cdot x} \phi_{x} \right|^{2} ,
  \label{eq:app-structurefactor}
\end{equation}
where the lattice momentum is $p = 2\pi n/L$.  $G(p)$ in Eq.\eqref{eq:app-structurefactor} is defined without subtraction of the disconnected piece, which vanishes identically at
$p \neq 0$ by translation invariance. 
The infrared diagnostic quoted in the tables is
the average over the two minimal momenta,
\begin{equation}
    G_p=\frac12\left[G(2\pi/L,0)+G(0,2\pi/L)\right]\,.
    \label{eq:Gp_min}
\end{equation}
The two are also
recorded separately since their difference
\begin{equation}
  \mathcal{A} = \frac{G(2\pi/L,0) - G(0,2\pi/L)}{G_{p}}
  \label{eq:app-asymmetry}
\end{equation}
is the rotational-isotropy check on an ensemble.
%

\subsection{Derived quantities}
\label{app:ops-derived}

The following observables are ratios or nonlinear functions of ensemble
averages of the operators defined above.  The susceptibility and Binder
cumulant are
\begin{equation}
  \chi = V\left(\langle m^2\rangle-\langle m\rangle^2\right),
  \qquad
  U_{4} = 1 - \frac{\langle m^{4}\rangle}
  {3\,\langle m^{2}\rangle^{2}} ,
  \label{eq:app-binder}
\end{equation}
where $\langle m\rangle=0$ by the $\mathbb{Z}_2$ symmetry of the finite-volume
ensemble.  The local kurtosis ratio is
\begin{equation}
  \frac{\langle\phi^{4}\rangle}{\langle\phi^{2}\rangle^{2}} .
\end{equation}

We use the second-moment correlation length,
\begin{equation}
  \xi =
  \frac{1}{2\sin(\pi/L)}
  \sqrt{\frac{\chi}{G_p}-1},
  \label{eq:app-xi}
\end{equation}
where $G_p$ is the correlation function at the smallest nonzero momentum
defined in Eq.~\eqref{eq:Gp_min}.  We frequently quote the dimensionless
ratio $\xi/L$.

For the finite-size scaling analysis we determine the exponent ratios $\gamma/\nu$ and $\beta/\nu$ from the
volume dependence
\begin{equation}
  \chi \propto L^{\gamma/\nu},
  \qquad
  \langle |m| \rangle \propto L^{-\beta/\nu}.
\end{equation}
The absolute value is used because the finite-volume ensemble preserves the
$\mathbb{Z}_2$ symmetry and hence $\langle m\rangle=0$.

The thermal exponent is obtained from derivatives with respect to $\kappa$.
For any observable $\mathcal O$,
\begin{equation}
  \frac{\partial \langle \mathcal{O} \rangle}{\partial \kappa}
    =
    2 V
    \left(
      \langle \mathcal{O}\,\mathrm{NN} \rangle
      -
      \langle \mathcal{O} \rangle \langle \mathrm{NN} \rangle
    \right),
  \label{eq:app-kappaderiv}
\end{equation}
since $-\partial S/\partial\kappa=2V\,\mathrm{NN}$ with the
axis-summed normalization of Eq.~\eqref{eq:app-nn}.  We use
\begin{equation}
  \frac{\partial}{\partial\kappa}\log\langle |m|\rangle,
  \qquad
  \frac{\partial}{\partial\kappa}\log\langle m^2\rangle,
  \qquad
  \frac{\partial U_4}{\partial\kappa},
\end{equation}
each of which scales as $L^{1/\nu}$ at criticality, up to corrections to scaling. We use these as independent estimators of $1/\nu$.

\subsection{Operator sets used in each role}
\label{app:ops-sets}

The $7\times7$ blocking kernel of Sec.~\ref{sec:perfect-kernel} is fitted using
\begin{equation}
\left\{
\begin{aligned}
&\phi^{2},\phi^{4},\phi^{6},
\mathrm{NN},\mathrm{2NN},\mathrm{diag},m^{2},\\
&G_{21},G_{22},G_{30},G_{31}
\end{aligned}
\right\}.
\end{equation}
The observables $m^4$, the local kurtosis
$\langle\phi^4\rangle/\langle\phi^2\rangle^2$, the action density, and
$G(p_{\min})$ are reserved for validation and are not included in the fit.

The conditional upscaling flow is trained by conditional maximum likelihood;
none of these observables enter its training loss.

Rethermalization is monitored using $S/V$, $\phi^2$, $\phi^4$,
$\mathrm{NN}$, $\mathrm{2NN}$, $\mathrm{diag}$, $m^2$, and $G(p_{\min})$,
together with the moments required to construct $\chi$, $U_4$, and $\xi/L$.

\subsection{Statistical treatment}
\label{app:statistics}

Linear observables are quoted as sample means with binned statistical errors.
Binning is used whenever measurements arise from a Markov chain, so that
autocorrelations are included in the uncertainty estimate.

Derived quantities are estimated by binned jackknife.  The estimator is
evaluated on the full sample and on each leave-one-bin-out subsample, and the
jackknife samples are propagated through the complete analysis.  When several
observables or fitted quantities enter a derived result, the same jackknife
deletions are used throughout so that correlations are preserved.

For cascade analyses, the statistically independent unit is the root
configuration, not an individual volume.  All descendants of a given root,
for example $L=32,64,\ldots,2048$, are resampled together.  This preserves the
cross-volume covariance generated by the recursive construction without
requiring an explicit model for it.  These correlations are substantial and
are included in all multivolume fits and derived quantities.

Distributional comparisons use the two-sample Kolmogorov--Smirnov (KS)
distance,
\begin{equation}
  D_{\rm KS}
  =
  \sup_x
  \left|F_{\rm direct}(x)-F_{\rm comparison}(x)\right|.
\end{equation}
Smaller values indicate greater overlap of the empirical cumulative
distributions. In addition, we consider the width ratio
\begin{equation}
    R_\sigma =
    \frac{\sigma_{\mathrm{blocked}}}{\sigma_{\mathrm{direct}}}.
\end{equation}
For independent ensembles, expectation values are compared using pulls with
uncertainties combined in quadrature. These distributional statistics are used as diagnostics rather than as the
sole figure of merit.  Marginal distributions can miss correlated or
collective discrepancies, which are probed separately by coupling-derivative
observables and finite-size-scaling tests.

\section{Monte Carlo renormalization-group analysis} 
\label{app:mcrg}

We briefly summarize the MCRG construction used in Sec.~\ref{sec:mcrg-test}, following Swendsen~\cite{Swendsen:1979gn}. Writing the blocked action after $n$ RG blocking steps as \begin{equation} 
S^{(n)} = \sum_\alpha K_\alpha^{(n)} S_\alpha^{(n)}, 
\end{equation}
the linearized RG transformation matrix  is 
\begin{equation} 
T_{\alpha\beta}^{(n)} = \frac{\partial K_\alpha^{(n+1)}} {\partial K_\beta^{(n)}}. 
\label{eq:mcrg-T} 
\end{equation}
The matrix $T^{(n)}$ is obtained directly from connected operator
correlations.
Defining 
\begin{align}
A_{\gamma\alpha}^{(n+1)} &= \left\langle S_\gamma^{(n+1)} S_\alpha^{(n+1)} \right\rangle_c , \\ 
B_{\gamma\beta}^{(n+1,n)} &= \left\langle S_\gamma^{(n+1)} S_\beta^{(n)} \right\rangle_c 
\end{align}
the chain rule gives 
\begin{equation}
B^{(n+1,n)} = A^{(n+1)} T^{(n)}. \label{eq:mcrg-ATB}
\end{equation}
Thus the linearized RG matrix can be determined without constructing the blocked action explicitly. We use the even operator basis 
\begin{align}
S_1 &= \sum_x \phi_x^2, & S_2 
&= \sum_x \phi_x^4, \\ S_3 &= \sum_{x,\mu} \phi_x\phi_{x+\hat\mu}, 
& S_4 &= \sum_x \phi_x\phi_{x+\hat x+\hat y}, \\ 
S_5 &= \sum_{x,\mu} \phi_x\phi_{x+2\hat\mu}, 
\end{align}
with periodic boundary conditions. 
We solve Eq.~\eqref{eq:mcrg-ATB} using an SVD rather than explicitly forming $A^{-1}$, and statistical uncertainties are determined by bootstrap resampling of the original configurations, with the full blocking hierarchy recomputed within each bootstrap sample. 
Near the fixed point, the eigenvalues of $T$ are related to the RG scaling
exponents by
\begin{equation}
\lambda_i = s^{y_i},
\end{equation}
for a scale-$s$ RG blocking transformation.  For $s=2$, the leading even
eigenvalue satisfies
\begin{equation}
\lambda_t = 2^{1/\nu},
\qquad
\nu = \frac{\log 2}{\log\lambda_t}.
\end{equation}

\section{Additional rethermalization tests}
\label{app:rethermalization}

Tables~\ref{tab:therm_l64_common_local} and~\ref{tab:therm_l64_common_long}  compare rethermalization  both with Wolff and HMC algorithms, and the deliberately mismatched critical-action test.  For the correctly matched critical source, Wolff and HMC updates show comparable rapid relaxation of the local observables.  The same behavior is seen for the source ensemble generated at criticality at  $\lambda_c=0.9$, $\kappa_c=0.342410$: despite a substantial sweep-zero mismatch in quantities such as the action density and $\langle\phi^4\rangle$, HMC restores the short-distance observables on essentially the same time scale as in the standard case.  By contrast, when the source coupling is deliberately mistuned away from criticality, the local observables again relax rapidly, but residual discrepancies persist much longer in infrared quantities such as $U_4$ and $\xi/L$.  This supports the RG interpretation that rethermalization efficiently removes short-distance and irrelevant-direction errors, while a mismatch with a component along the relevant thermal direction is much more difficult to repair.



\begin{table*}[t]
\centering
\caption{L32 $\to$ L64 rethermalization. Wolff I/II and HMC I/II use $\lambda_c=1$ source couplings $\kappa_c=0.340301$ and $0.340100$; HMC III uses $\lambda_c=0.9$, $\kappa_c=0.342410$. HMC I--III target $\lambda_f=1$, $\kappa_f=0.340301$, whereas HMC IV deliberately targets $\kappa_f=0.340100$. HMC II and III combine five independent replicas at every displayed sweep ($N=10000$). The two direct-native columns match their respective fine targets and are shown only at sweep zero. Errors are configuration-bootstrap standard deviations.}
\scriptsize
\begin{ruledtabular}
\begin{tabular}{lcccc|ccc|cc}
Observable & Native $\kappa=.340301$ & Sweep & Wolff I & Wolff II & HMC I & HMC II & HMC III & Native $\kappa=.340100$ & HMC IV \\
\hline
$S/V$ & -0.55001(23) & 0 & -0.52684(42) & -0.52544(56) & -0.52699(30) & -0.52603(31) & -0.52107(30) & -0.54743(38) & -0.52617(76) \\
 &  & 10 & -0.55026(38) & -0.54973(47) & -0.54868(27) & -0.54762(26) & -0.54779(26) &  & -0.54681(68) \\
 &  & 20 & -0.55007(37) & -0.55051(48) & -0.54972(26) & -0.54896(27) & -0.54910(27) &  & -0.54808(66) \\
 &  & 50 & -0.54997(38) & -0.55007(48) & -0.54984(26) & -0.54956(27) & -0.54951(27) &  & -0.54757(69) \\
 &  & 100 & -0.54983(36) & -0.55152(49) & -0.55000(26) & -0.54934(26) & -0.54971(26) &  & -0.54776(65) \\
 &  & 250 &  &  & -0.54980(27) & -0.54960(27) & -0.54969(27) &  & -0.54837(70) \\
 &  & 500 &  &  & -0.55011(27) & -0.54969(27) & -0.55051(26) &  & -0.54783(70) \\
 &  & 750 &  &  & -0.54976(27) & -0.55029(26) & -0.55007(27) &  & -0.54769(67) \\
 &  & 1000 &  &  & -0.55029(27) & -0.54988(26) & -0.54967(26) &  & -0.54772(68) \\
 &  & 1250 &  &  & -0.55013(26) & -0.54977(27) & -0.55005(27) &  & -0.54719(67) \\
 &  & 1500 &  &  & -0.54992(27) & -0.54973(26) & -0.54958(27) &  & -0.54756(70) \\
\hline
$\langle\phi^2\rangle$ & 0.82832(15) & 0 & 0.82988(23) & 0.82902(30) & 0.83002(16) & 0.82924(16) & 0.83122(17) & 0.82675(24) & 0.82981(43) \\
 &  & 10 & 0.82856(24) & 0.82797(30) & 0.82767(17) & 0.82717(17) & 0.82730(17) &  & 0.82721(45) \\
 &  & 20 & 0.82843(24) & 0.82840(31) & 0.82801(17) & 0.82777(17) & 0.82780(17) &  & 0.82732(43) \\
 &  & 50 & 0.82847(24) & 0.82848(31) & 0.82810(17) & 0.82809(17) & 0.82795(17) &  & 0.82762(44) \\
 &  & 100 & 0.82820(24) & 0.82907(32) & 0.82838(17) & 0.82810(17) & 0.82799(17) &  & 0.82718(43) \\
 &  & 250 &  &  & 0.82806(17) & 0.82805(17) & 0.82798(17) &  & 0.82808(44) \\
 &  & 500 &  &  & 0.82819(17) & 0.82816(17) & 0.82862(17) &  & 0.82726(45) \\
 &  & 750 &  &  & 0.82817(17) & 0.82838(17) & 0.82841(17) &  & 0.82716(43) \\
 &  & 1000 &  &  & 0.82851(17) & 0.82832(17) & 0.82825(17) &  & 0.82673(43) \\
 &  & 1250 &  &  & 0.82838(17) & 0.82797(17) & 0.82838(17) &  & 0.82697(44) \\
 &  & 1500 &  &  & 0.82822(17) & 0.82831(17) & 0.82821(17) &  & 0.82705(43) \\
\hline
$\langle\phi^4\rangle$ & 1.05001(30) & 0 & 1.07759(43) & 1.07640(57) & 1.07791(31) & 1.07655(31) & 1.08579(32) & 1.04718(47) & 1.07743(81) \\
 &  & 10 & 1.05067(46) & 1.04944(59) & 1.04907(33) & 1.04834(32) & 1.04845(33) &  & 1.04857(87) \\
 &  & 20 & 1.05045(46) & 1.05003(60) & 1.04945(33) & 1.04901(33) & 1.04907(33) &  & 1.04822(84) \\
 &  & 50 & 1.05059(46) & 1.05053(60) & 1.04962(33) & 1.04969(33) & 1.04947(33) &  & 1.04923(86) \\
 &  & 100 & 1.04992(46) & 1.05132(62) & 1.05017(34) & 1.04978(32) & 1.04938(33) &  & 1.04810(84) \\
 &  & 250 &  &  & 1.04958(33) & 1.04962(34) & 1.04941(33) &  & 1.04951(86) \\
 &  & 500 &  &  & 1.04979(33) & 1.04978(33) & 1.05064(33) &  & 1.04811(86) \\
 &  & 750 &  &  & 1.04981(33) & 1.05019(34) & 1.05022(34) &  & 1.04806(84) \\
 &  & 1000 &  &  & 1.05045(33) & 1.05024(33) & 1.05014(32) &  & 1.04706(84) \\
 &  & 1250 &  &  & 1.05013(33) & 1.04939(33) & 1.05018(33) &  & 1.04769(85) \\
 &  & 1500 &  &  & 1.05000(33) & 1.05026(33) & 1.05010(33) &  & 1.04761(84) \\
\hline
$\mathrm{NN}$ & 1.13386(50) & 0 & 1.13805(79) & 1.1355(10) & 1.13851(56) & 1.13626(57) & 1.13964(58) & 1.12887(79) & 1.1376(15) \\
 &  & 10 & 1.13483(79) & 1.13312(100) & 1.13148(56) & 1.12957(55) & 1.12981(55) &  & 1.1293(15) \\
 &  & 20 & 1.13442(78) & 1.1345(10) & 1.13306(55) & 1.13165(56) & 1.13190(57) &  & 1.1305(14) \\
 &  & 50 & 1.13443(79) & 1.1345(10) & 1.13335(55) & 1.13306(56) & 1.13287(56) &  & 1.1308(15) \\
 &  & 100 & 1.13364(77) & 1.1369(10) & 1.13399(56) & 1.13284(55) & 1.13297(56) &  & 1.1301(14) \\
 &  & 250 &  &  & 1.13328(56) & 1.13307(57) & 1.13299(56) &  & 1.1317(15) \\
 &  & 500 &  &  & 1.13385(56) & 1.13328(57) & 1.13505(55) &  & 1.1301(15) \\
 &  & 750 &  &  & 1.13340(57) & 1.13444(56) & 1.13411(57) &  & 1.1299(14) \\
 &  & 1000 &  &  & 1.13462(57) & 1.13400(56) & 1.13364(55) &  & 1.1292(14) \\
 &  & 1250 &  &  & 1.13411(55) & 1.13310(56) & 1.13407(56) &  & 1.1289(14) \\
 &  & 1500 &  &  & 1.13384(56) & 1.13381(55) & 1.13353(56) &  & 1.1293(15) \\
\hline
$2\mathrm{NN}$ & 0.94918(68) & 0 & 0.9569(11) & 0.9526(15) & 0.95701(78) & 0.95358(80) & 0.95823(80) & 0.9423(11) & 0.9566(20) \\
 &  & 10 & 0.9508(11) & 0.9484(14) & 0.94619(77) & 0.94358(77) & 0.94380(76) &  & 0.9433(20) \\
 &  & 20 & 0.9497(11) & 0.9502(14) & 0.94852(76) & 0.94610(78) & 0.94670(79) &  & 0.9452(19) \\
 &  & 50 & 0.9502(11) & 0.9505(14) & 0.94880(76) & 0.94823(78) & 0.94798(77) &  & 0.9452(20) \\
 &  & 100 & 0.9494(11) & 0.9538(14) & 0.94951(78) & 0.94800(77) & 0.94811(77) &  & 0.9446(19) \\
 &  & 250 &  &  & 0.94884(77) & 0.94828(78) & 0.94839(78) &  & 0.9464(20) \\
 &  & 500 &  &  & 0.94969(78) & 0.94853(78) & 0.95088(75) &  & 0.9452(20) \\
 &  & 750 &  &  & 0.94879(78) & 0.95029(77) & 0.94966(78) &  & 0.9439(20) \\
 &  & 1000 &  &  & 0.95039(78) & 0.94978(77) & 0.94917(76) &  & 0.9432(20) \\
 &  & 1250 &  &  & 0.94967(77) & 0.94851(78) & 0.94986(77) &  & 0.9428(19) \\
 &  & 1500 &  &  & 0.94928(77) & 0.94934(76) & 0.94898(78) &  & 0.9430(20) \\
\hline
$\mathrm{diag}$ & 0.50786(30) & 0 & 0.51314(48) & 0.51135(64) & 0.51331(34) & 0.51185(35) & 0.51369(35) & 0.50472(48) & 0.51294(88) \\
 &  & 10 & 0.50854(48) & 0.50748(60) & 0.50651(34) & 0.50527(34) & 0.50540(33) &  & 0.50517(87) \\
 &  & 20 & 0.50810(47) & 0.50823(61) & 0.50755(33) & 0.50643(34) & 0.50658(35) &  & 0.50596(86) \\
 &  & 50 & 0.50813(48) & 0.50834(62) & 0.50766(33) & 0.50742(34) & 0.50729(34) &  & 0.50602(87) \\
 &  & 100 & 0.50786(46) & 0.50967(63) & 0.50786(34) & 0.50733(34) & 0.50742(34) &  & 0.50591(83) \\
 &  & 250 &  &  & 0.50763(34) & 0.50745(34) & 0.50750(34) &  & 0.50639(88) \\
 &  & 500 &  &  & 0.50795(34) & 0.50748(34) & 0.50859(33) &  & 0.50597(87) \\
 &  & 750 &  &  & 0.50767(34) & 0.50831(34) & 0.50785(34) &  & 0.50568(85) \\
 &  & 1000 &  &  & 0.50829(34) & 0.50798(34) & 0.50779(33) &  & 0.50531(86) \\
 &  & 1250 &  &  & 0.50794(33) & 0.50746(34) & 0.50806(34) &  & 0.50510(85) \\
 &  & 1500 &  &  & 0.50792(34) & 0.50798(33) & 0.50769(34) &  & 0.50505(88) \\
\hline
\end{tabular}
\end{ruledtabular}
\label{tab:therm_l64_common_local}
\end{table*}

\begin{table*}[t]
\centering
\caption{L32 $\to$ L64 rethermalization. Wolff I/II and HMC I/II use $\lambda_c=1$ source couplings $\kappa_c=0.340301$ and $0.340100$; HMC III uses $\lambda_c=0.9$, $\kappa_c=0.342410$. HMC I--III target $\lambda_f=1$, $\kappa_f=0.340301$, whereas HMC IV deliberately targets $\kappa_f=0.340100$. HMC II and III combine five independent replicas at every displayed sweep ($N=10000$). The two direct-native columns match their respective fine targets and are shown only at sweep zero. Errors are configuration-bootstrap standard deviations.}
\scriptsize
\begin{ruledtabular}
\begin{tabular}{lcccc|ccc|cc}
Observable & Native $\kappa=.340301$ & Sweep & Wolff I & Wolff II & HMC I & HMC II & HMC III & Native $\kappa=.340100$ & HMC IV \\
\hline
$\chi$ & 1223.1(46) & 0 & 1244.0(72) & 1211.4(97) & 1242.5(51) & 1223.7(52) & 1244.3(51) & 1189.9(74) & 1248(13) \\
 &  & 10 & 1238.9(71) & 1234.6(91) & 1224.4(51) & 1208.3(51) & 1217.9(49) &  & 1225(13) \\
 &  & 20 & 1228.7(71) & 1238.2(93) & 1226.6(50) & 1213.2(51) & 1223.9(52) &  & 1227(13) \\
 &  & 50 & 1232.8(72) & 1233.6(92) & 1226.4(51) & 1218.1(52) & 1226.0(50) &  & 1237(13) \\
 &  & 100 & 1231.9(71) & 1259.4(93) & 1228.1(51) & 1218.7(50) & 1227.5(51) &  & 1233(13) \\
 &  & 250 &  &  & 1232.1(51) & 1223.5(52) & 1230.3(52) &  & 1236(13) \\
 &  & 500 &  &  & 1236.0(51) & 1223.9(51) & 1233.9(51) &  & 1226(13) \\
 &  & 750 &  &  & 1230.8(51) & 1233.8(51) & 1230.3(52) &  & 1209(13) \\
 &  & 1000 &  &  & 1234.4(51) & 1234.6(52) & 1230.3(50) &  & 1207(13) \\
 &  & 1250 &  &  & 1233.7(51) & 1230.3(50) & 1233.4(51) &  & 1192(13) \\
 &  & 1500 &  &  & 1234.4(51) & 1229.3(50) & 1234.3(51) &  & 1192(14) \\
\hline
$U_4$ & 0.60820(97) & 0 & 0.6100(15) & 0.6033(21) & 0.6087(11) & 0.6067(11) & 0.6101(11) & 0.6024(17) & 0.6135(25) \\
 &  & 10 & 0.6121(14) & 0.6116(19) & 0.6090(11) & 0.6072(11) & 0.6103(11) &  & 0.6138(25) \\
 &  & 20 & 0.6095(15) & 0.6105(20) & 0.6091(11) & 0.6074(11) & 0.6107(11) &  & 0.6144(25) \\
 &  & 50 & 0.6101(15) & 0.6101(19) & 0.6094(11) & 0.6078(11) & 0.6106(11) &  & 0.6142(25) \\
 &  & 100 & 0.6106(15) & 0.6136(18) & 0.6084(11) & 0.6081(11) & 0.6099(11) &  & 0.6155(25) \\
 &  & 250 &  &  & 0.6103(11) & 0.6080(11) & 0.6101(11) &  & 0.6140(25) \\
 &  & 500 &  &  & 0.6106(10) & 0.6081(11) & 0.6099(11) &  & 0.6106(27) \\
 &  & 750 &  &  & 0.6105(11) & 0.6097(11) & 0.6087(11) &  & 0.6103(28) \\
 &  & 1000 &  &  & 0.6102(10) & 0.6102(11) & 0.6093(11) &  & 0.6086(27) \\
 &  & 1250 &  &  & 0.6095(11) & 0.6102(11) & 0.6102(11) &  & 0.6064(28) \\
 &  & 1500 &  &  & 0.6099(11) & 0.6101(11) & 0.6092(11) &  & 0.6037(30) \\
\hline
$\xi/L$ & 0.8842(68) & 0 & 0.898(11) & 0.868(14) & 0.8953(78) & 0.8821(76) & 0.9021(79) & 0.846(10) & 0.924(20) \\
 &  & 10 & 0.905(11) & 0.903(14) & 0.8948(78) & 0.8830(76) & 0.8977(77) &  & 0.927(20) \\
 &  & 20 & 0.892(11) & 0.896(15) & 0.8908(75) & 0.8841(77) & 0.9015(79) &  & 0.924(21) \\
 &  & 50 & 0.895(11) & 0.896(14) & 0.8922(78) & 0.8842(78) & 0.9006(78) &  & 0.941(21) \\
 &  & 100 & 0.900(11) & 0.932(15) & 0.8908(78) & 0.8848(76) & 0.8970(78) &  & 0.935(21) \\
 &  & 250 &  &  & 0.9012(77) & 0.8874(79) & 0.8983(78) &  & 0.935(21) \\
 &  & 500 &  &  & 0.9087(79) & 0.8878(77) & 0.8934(78) &  & 0.915(20) \\
 &  & 750 &  &  & 0.9007(79) & 0.8987(78) & 0.8961(78) &  & 0.895(20) \\
 &  & 1000 &  &  & 0.8994(78) & 0.9072(80) & 0.8962(77) &  & 0.880(20) \\
 &  & 1250 &  &  & 0.9016(79) & 0.9042(78) & 0.9043(78) &  & 0.855(19) \\
 &  & 1500 &  &  & 0.9045(79) & 0.8975(77) & 0.9071(78) &  & 0.861(20) \\
\hline
\end{tabular}
\end{ruledtabular}
\label{tab:therm_l64_common_long}
\end{table*}

\section{Additional plots and tables}
\label{app:extra-plots}

  \begin{table}[t]
  \centering
  \caption{
Perfect-blocking comparison on the $L=32$ coarse lattice:
direct $L=32$ configurations versus $L=64$ configurations blocked to
$L=32$.  The final columns give the two-sample KS distance and the ratio $\sigma_{\rm blocked}/\sigma_{\rm direct}$.
  }
  \label{tab:perfect_blocking_L32}
  \begin{ruledtabular}
  \begin{tabular}{lcccc}
  Operator & Direct $L=32$ & Blocked $64\to32$ & KS & Std. ratio \\
  \hline
  $S/V$ & $-0.55454(49)$ & $-0.55517(60)$ & $0.0439$ & $0.867$ \\
  $\phi^2$ & $0.83090(32)$ & $0.82903(51)$ & $0.0471$ & $1.121$ \\
  $\phi^4$ & $1.05487(64)$ & $1.0501(11)$ & $0.0592$ & $1.165$ \\
  $\langle\phi^4\rangle/\langle\phi^2\rangle^2$
   & $1.52791(36)$ & $1.52746(49)$ & $0.0211$ & $0.954$ \\
  $\mathrm{NN}$ & $1.1439(10)$ & $1.1406(15)$ & $0.0197$ & $1.018$ \\
  $2\mathrm{nn}$ & $0.9646(14)$ & $0.9615(20)$ & $0.0188$ & $0.985$ \\
  $\mathrm{diag}$ & $0.51417(63)$ & $0.51218(87)$ & $0.0249$ & $0.986$ \\
  $m^2$ & $0.3582(15)$ & $0.3540(21)$ & $0.0204$ & $1.000$ \\
  $m^4$ & $0.15089(97)$ & $0.1478(14)$ & $0.0204$ & $0.989$ \\
  $G(p_{\min})$ & $11.36(15)$ & $11.67(22)$ & $0.0180$ & $1.011$ \\
  \end{tabular}
  \end{ruledtabular}
  \end{table}

  \begin{table}[t]
  \centering
  \caption{
Perfect-blocking comparison on the $L=64$ coarse lattice:
direct $L=64$ configurations versus $L=128$ configurations blocked to
$L=64$.  The final columns give the two-sample KS distance and the ratio $\sigma_{\rm blocked}/\sigma_{\rm direct}$.
  }
  \label{tab:perfect_blocking_L64}
  \begin{ruledtabular}
  \begin{tabular}{lcccc}
  Operator & Direct $L=64$ & Blocked $128\to64$ & KS & Std. ratio \\
  \hline
  $S/V$ & $-0.54991(38)$ & $-0.55206(32)$ & $0.0718$ & $0.850$ \\
  $\phi^2$ & $0.82835(24)$ & $0.82600(27)$ & $0.0658$ & $1.122$ \\
  $\phi^4$ & $1.05010(47)$ & $1.04432(55)$ & $0.0854$ & $1.176$ \\
  $\langle\phi^4\rangle/\langle\phi^2\rangle^2$
   & $1.53041(27)$ & $1.53052(26)$ & $0.0210$ & $0.953$ \\
  $\mathrm{NN}$ & $1.1338(8)$ & $1.1319(8)$ & $0.0166$ & $1.010$ \\
  $2\mathrm{nn}$ & $0.9489(11)$ & $0.9489(11)$ & $0.0098$ & $0.984$ \\
  $\mathrm{diag}$ & $0.50773(48)$ & $0.50714(47)$ & $0.0118$ & $0.986$ \\
  $m^2$ & $0.2980(18)$ & $0.2994(18)$ & $0.0230$ & $1.000$ \\
  $m^4$ & $0.10458(95)$ & $0.10540(94)$ & $0.0230$ & $0.996$ \\
  $G(p_{\min})$ & $38.61(72)$ & $38.37(72)$ & $0.0142$ & $0.992$ \\
  \end{tabular}
  \end{ruledtabular}
  \end{table}

\begin{figure*}
      \centering
      \includegraphics[width=0.3\linewidth]{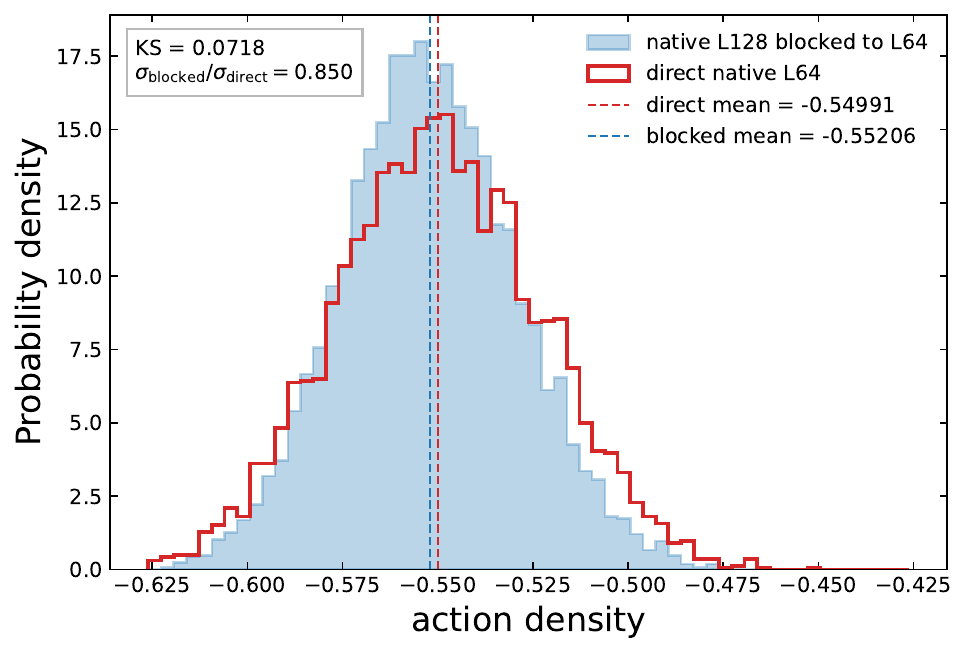} 
      \includegraphics[width=0.3\linewidth]{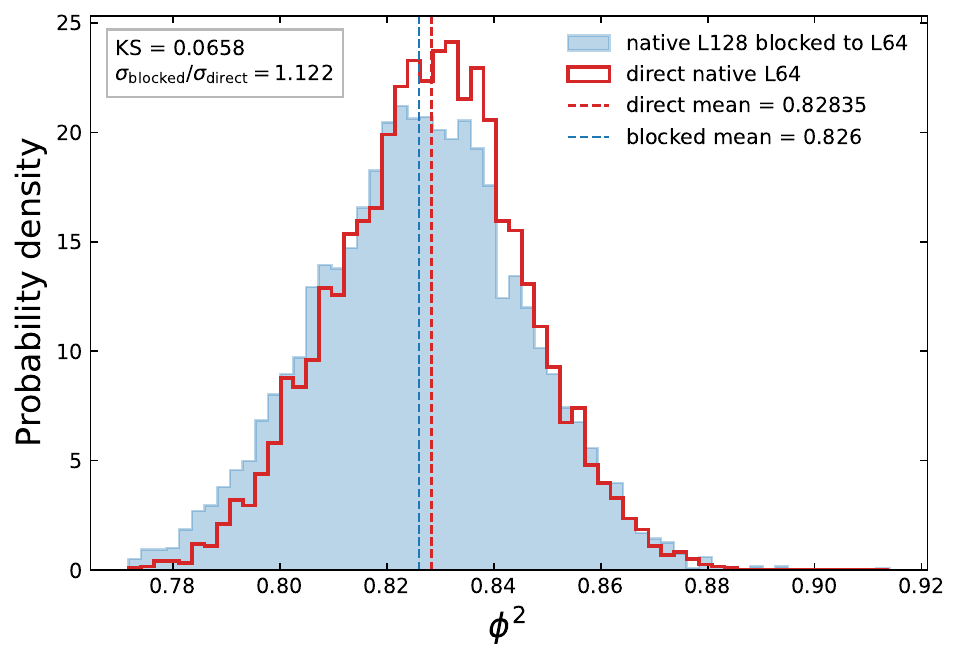} 
      \includegraphics[width=0.3\linewidth]{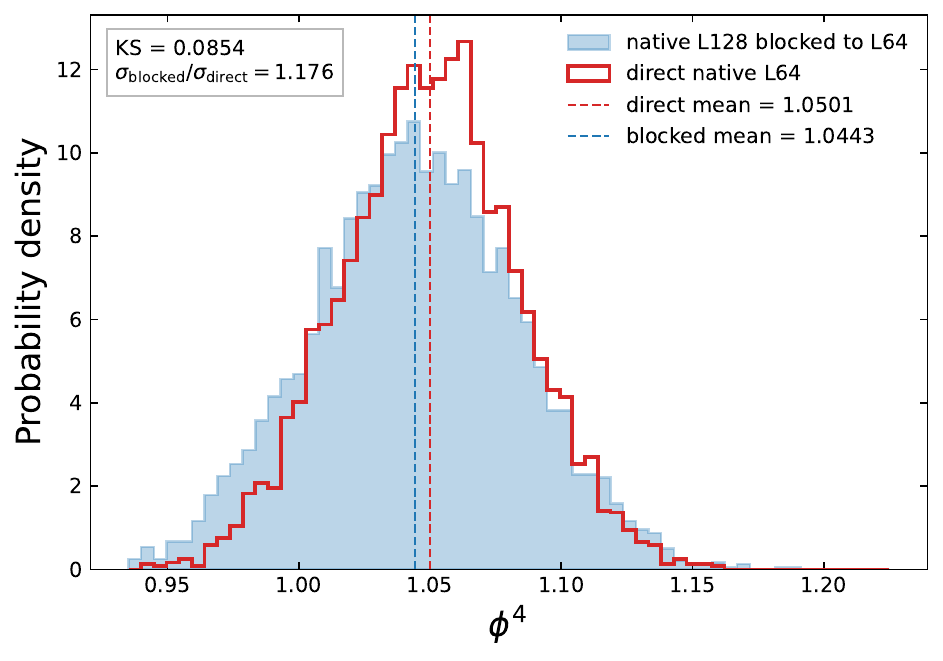} 
      \includegraphics[width=0.3\linewidth]{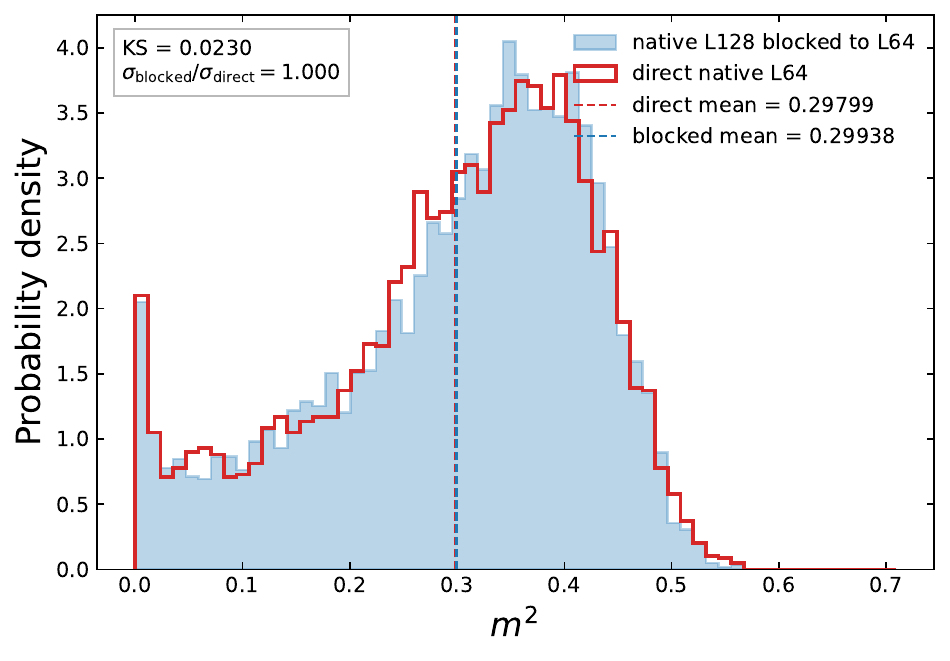}
      \includegraphics[width=0.3\linewidth]{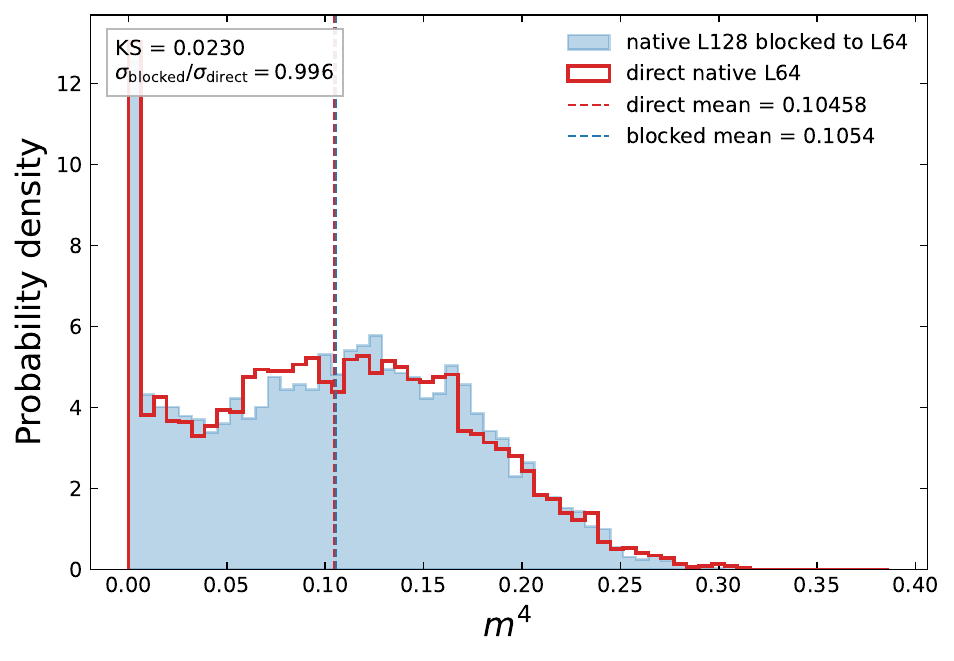}
      \includegraphics[width=0.3\linewidth]{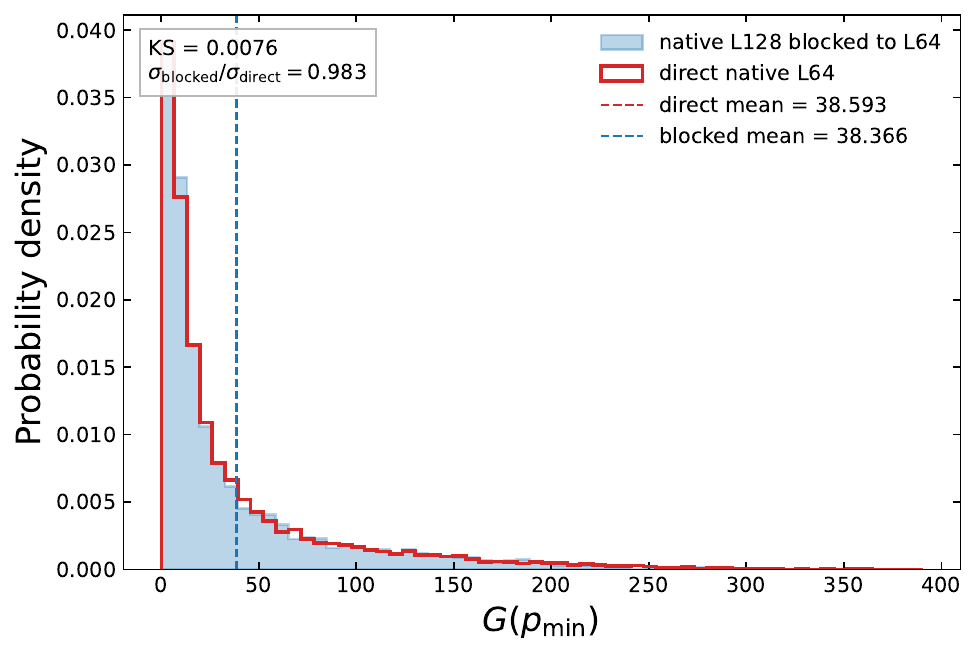}
      \caption{Perfect-blocking test for $L=128\to64$, using the same kernel optimized
from $L=32\to16$ blocking.  The panels show the same observables as
Fig.~\ref{fig:L32-to-16-blocking-histograms}.}
      \label{fig:L128-to-64-blocking-histograms}
\end{figure*}

\begin{figure*}
      \centering
      \includegraphics[width=0.3\linewidth]{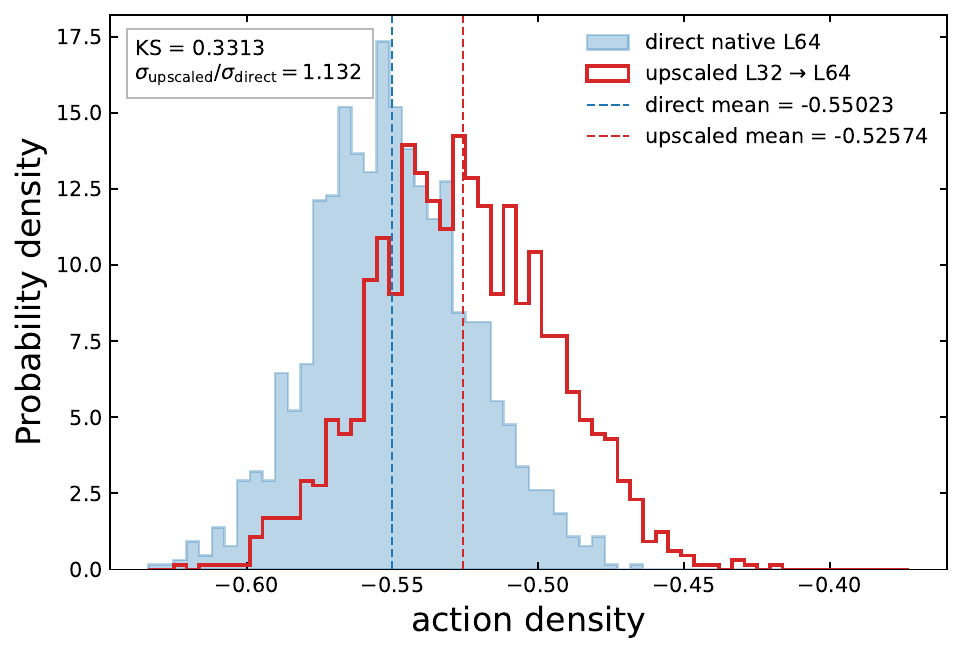} 
      \includegraphics[width=0.3\linewidth]{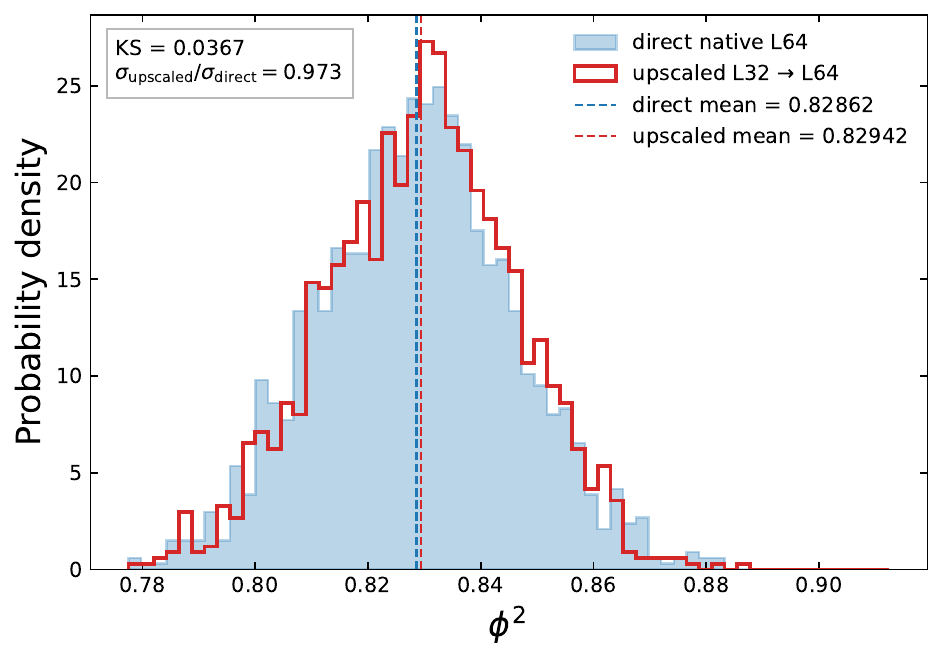} 
      \includegraphics[width=0.3\linewidth]{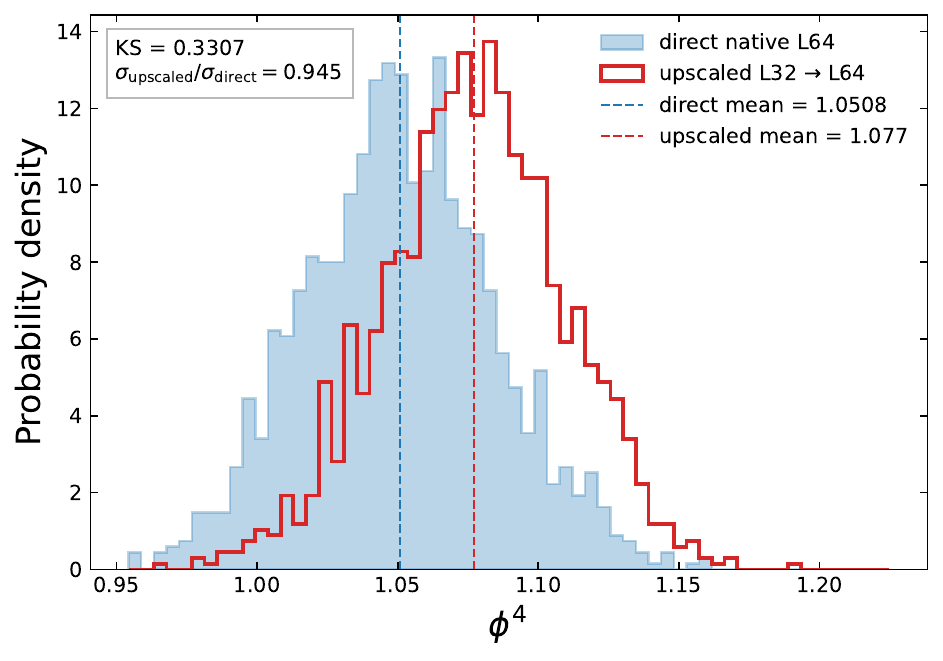} 
      \includegraphics[width=0.3\linewidth]{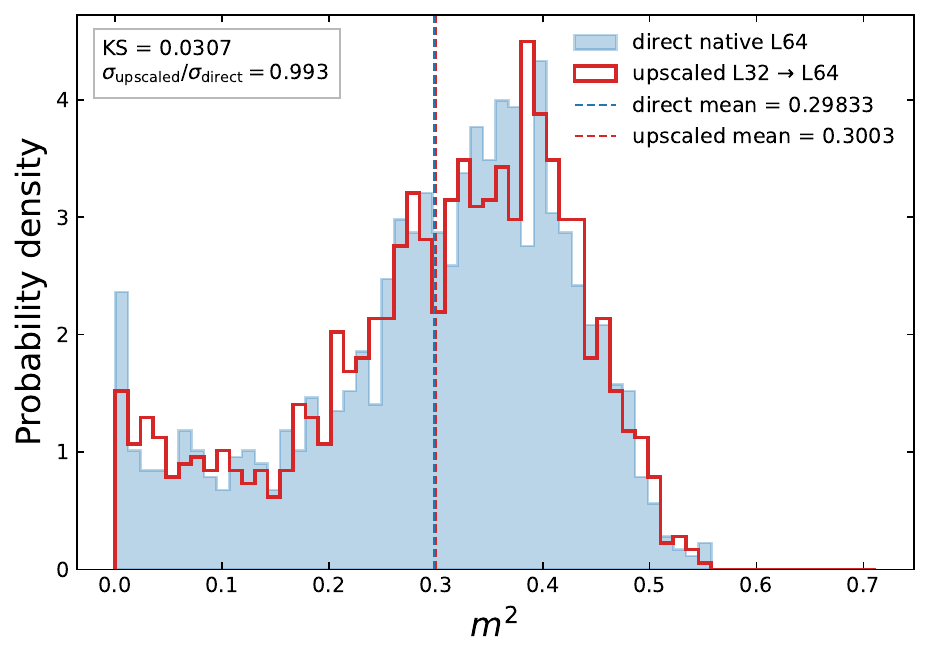}
      \includegraphics[width=0.3\linewidth]{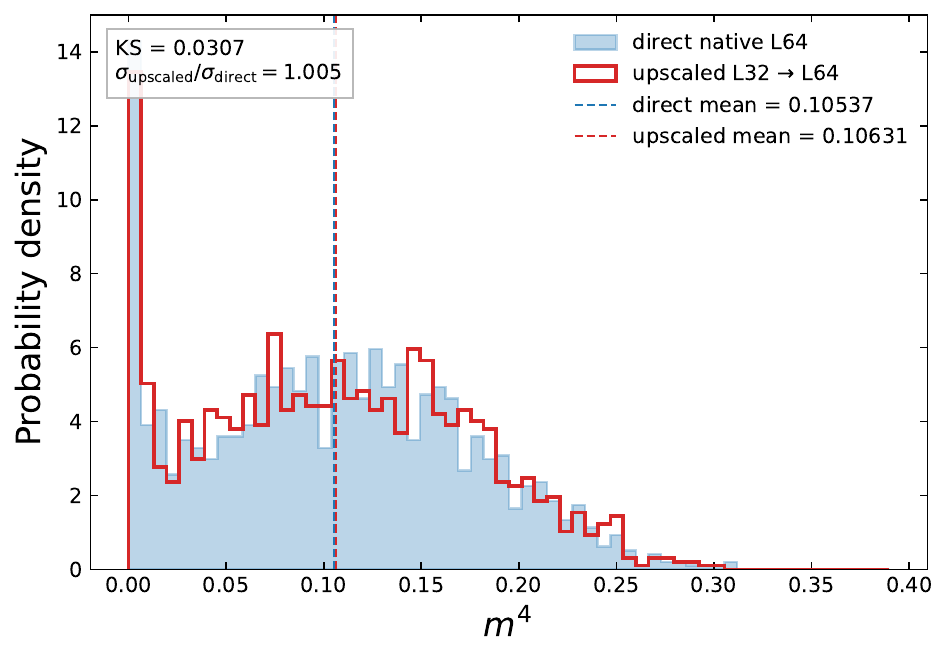}
      \includegraphics[width=0.3\linewidth]{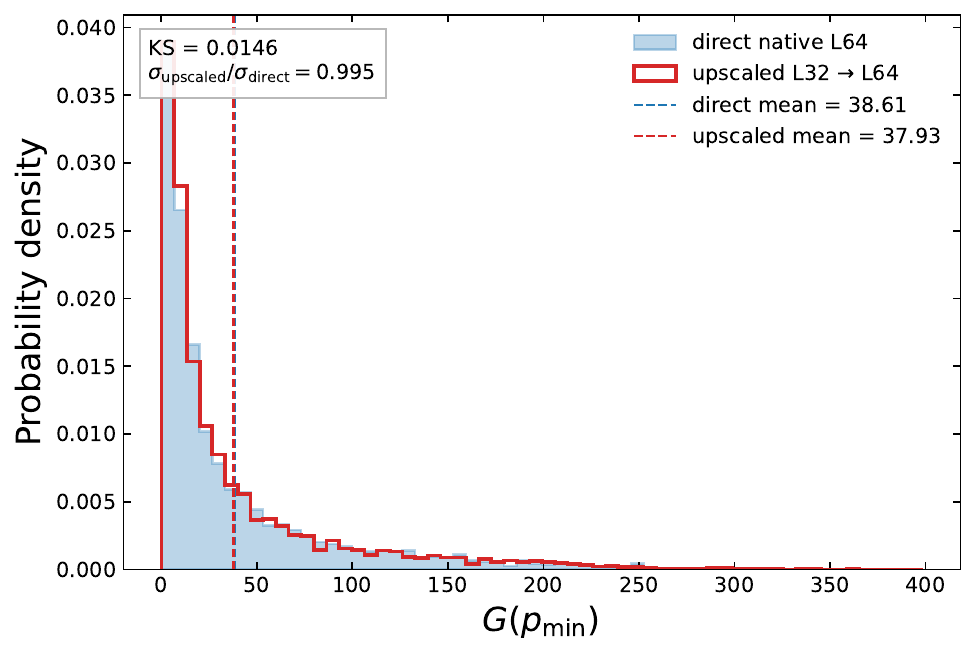}
      \caption{Raw conditional-flow proposal for $L=32\to64$, using the flow trained on $L=16\to32$ without retraining.  The panels show the same observables as Fig.~\ref{fig:L16-to-32-initial-upscale}.  Local discrepancies increase, while the long-distance distributions remain comparatively stable.
}
      \label{fig:L32-to-64-initial-upscale}
\end{figure*}

\begin{figure*}
      \centering
      \includegraphics[width=0.3\linewidth]{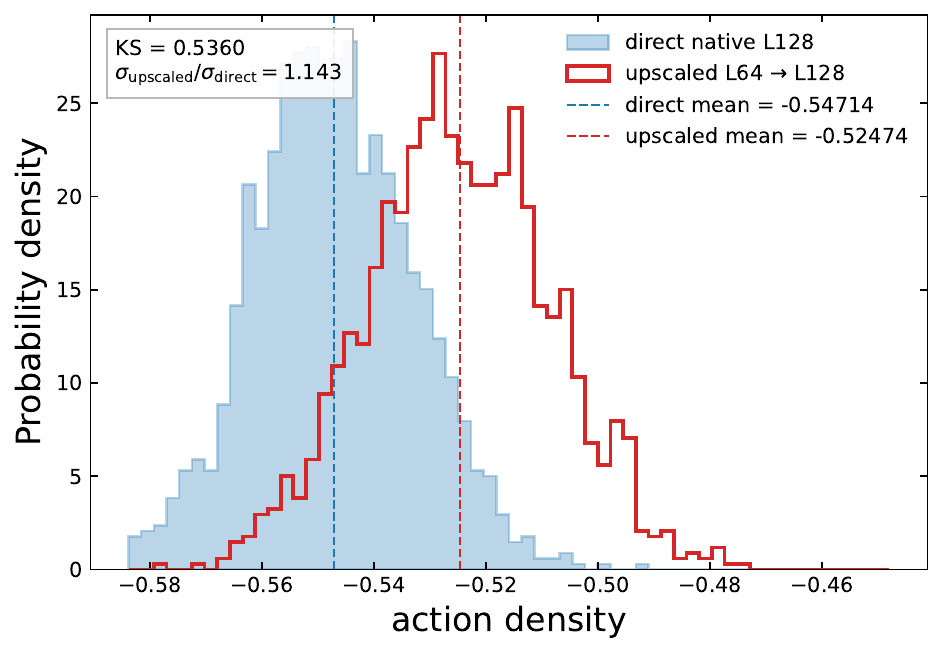} 
      \includegraphics[width=0.3\linewidth]{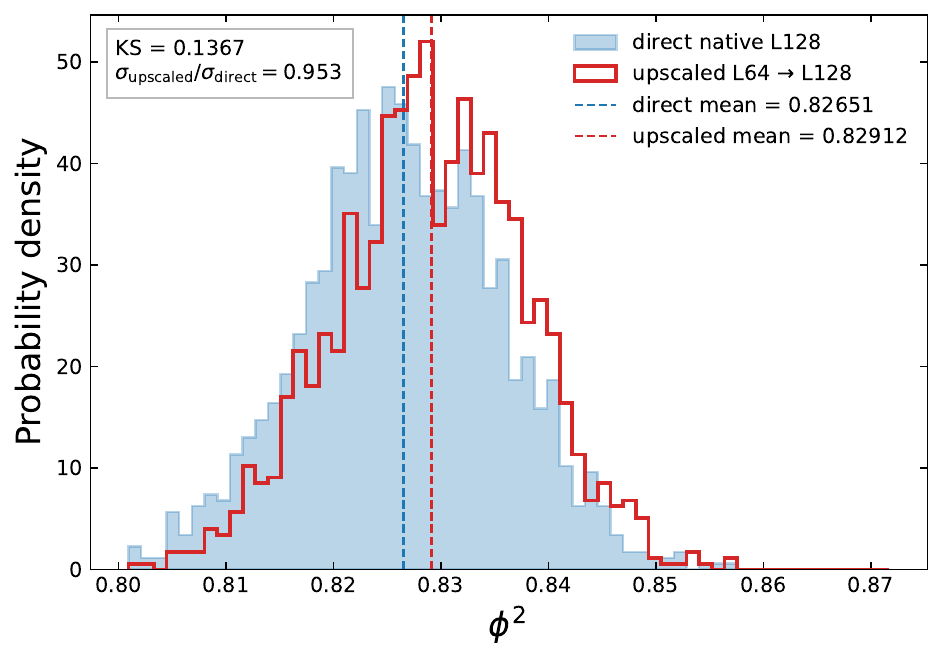} 
      \includegraphics[width=0.3\linewidth]{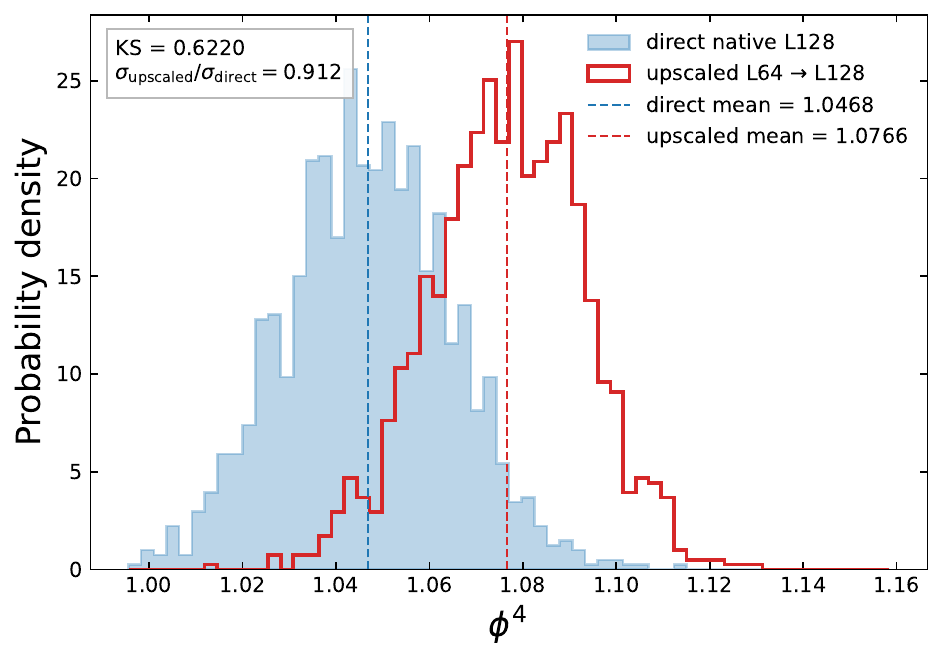} 
      \includegraphics[width=0.3\linewidth]{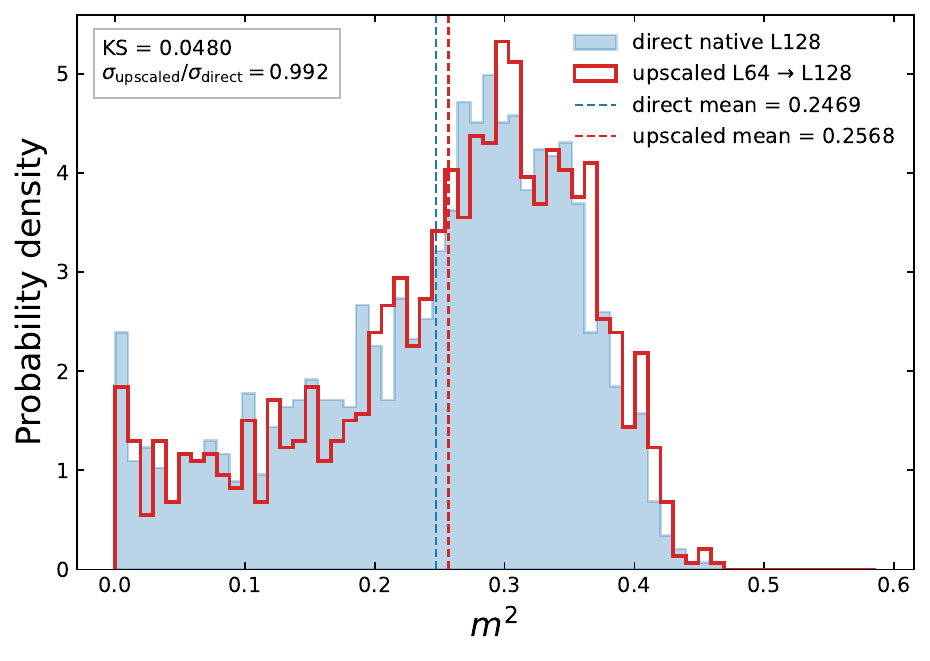}
      \includegraphics[width=0.3\linewidth]{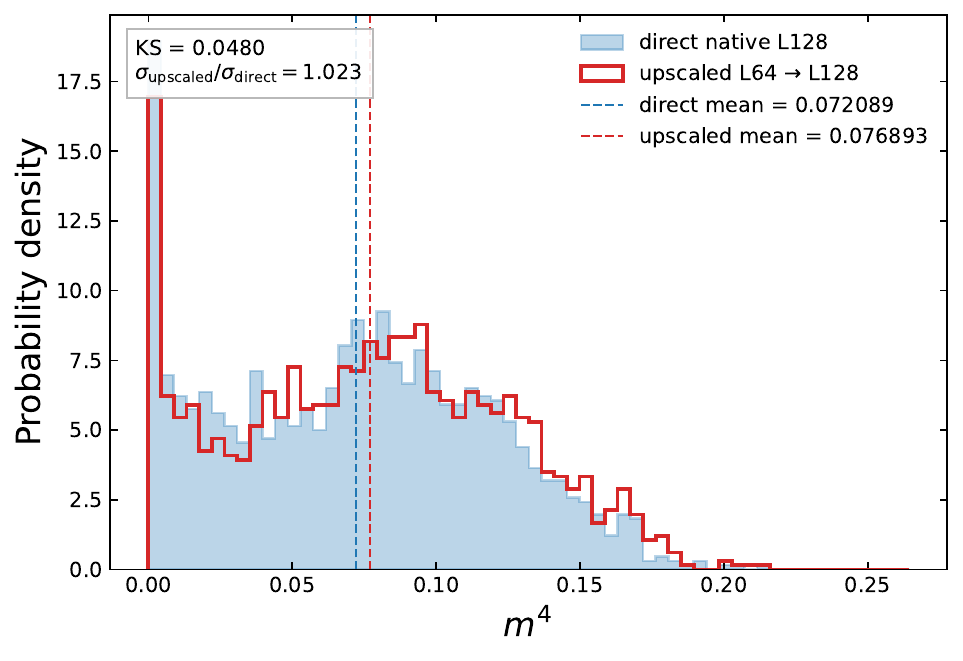}
      \includegraphics[width=0.3\linewidth]{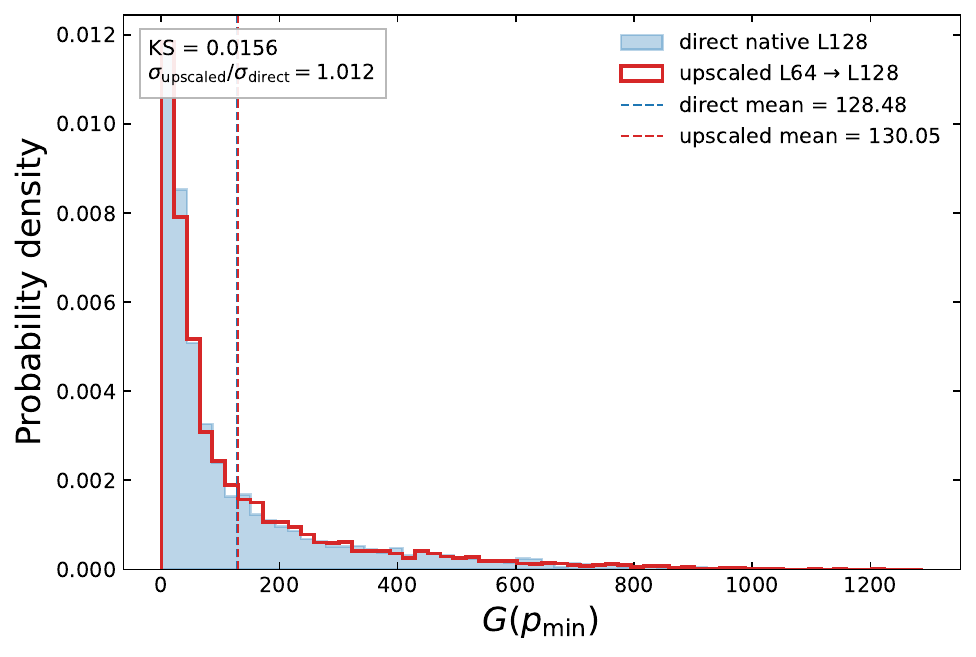}
      \caption{Raw conditional-flow proposal for $L=64\to128$, using the flow trained on $L=16\to32$ without retraining.  The panels show the same observables as Fig.~\ref{fig:L16-to-32-initial-upscale}.  The increasing local mismatch illustrates the finite-volume transfer error of the raw flow proposal.
}
      \label{fig:L64-to-128-initial-upscale}
\end{figure*}

\begin{figure*}
      \centering
      \includegraphics[width=0.3\linewidth]{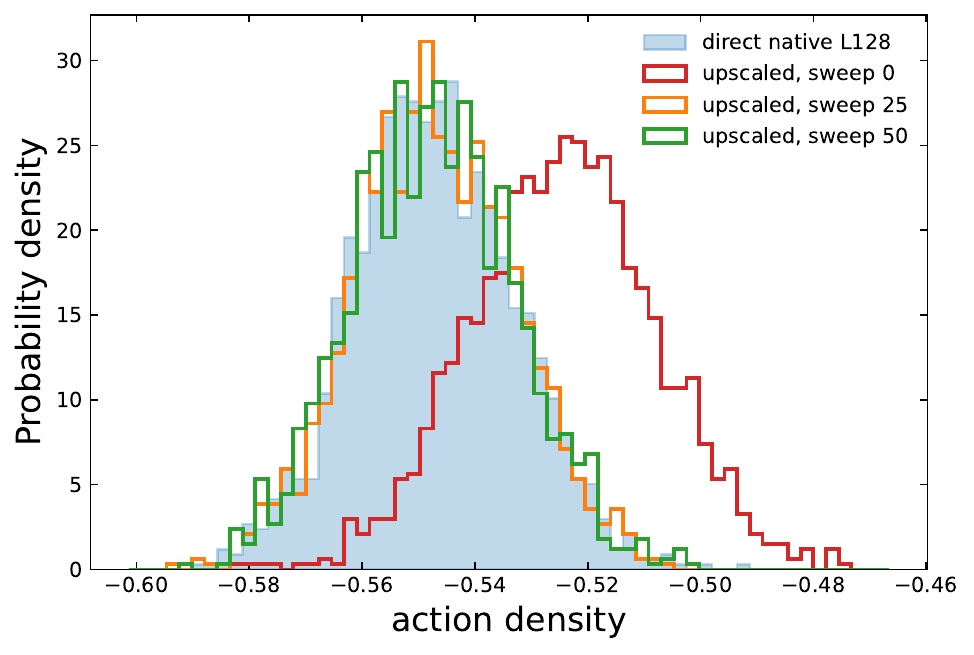} 
      \includegraphics[width=0.3\linewidth]{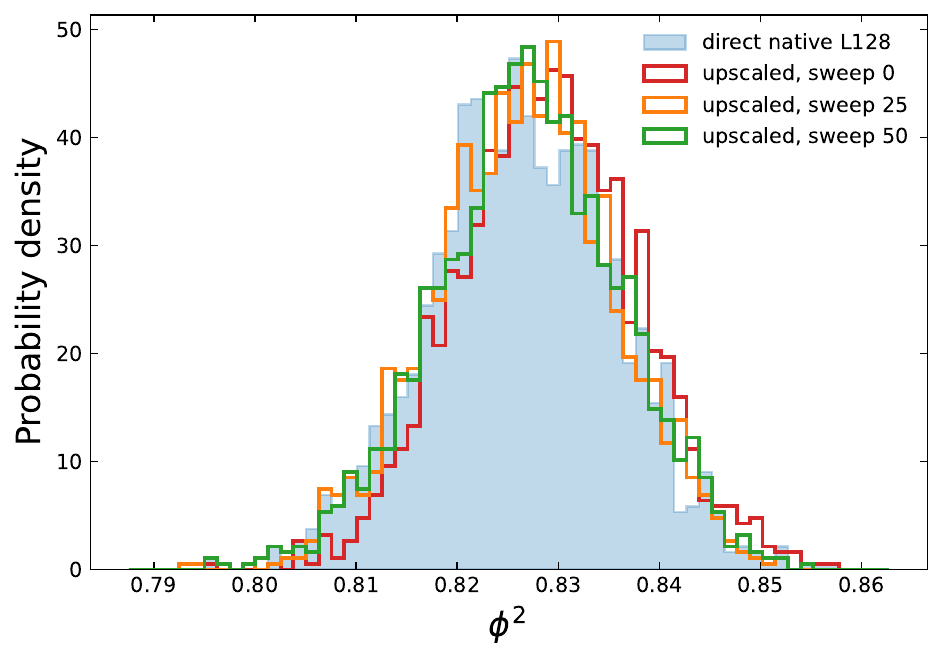} 
      \includegraphics[width=0.3\linewidth]{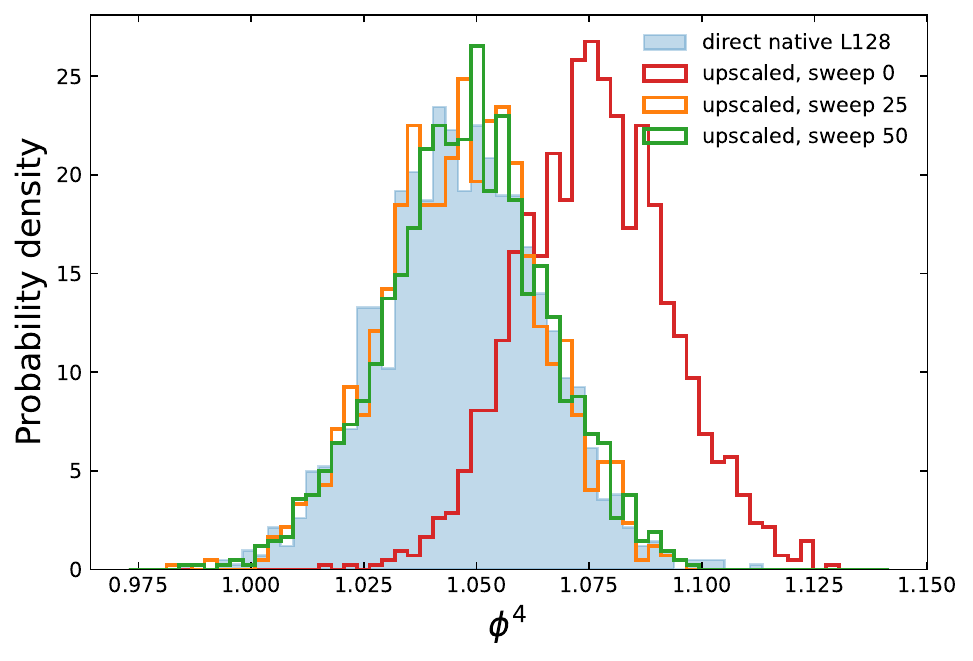} 
      \includegraphics[width=0.3\linewidth]{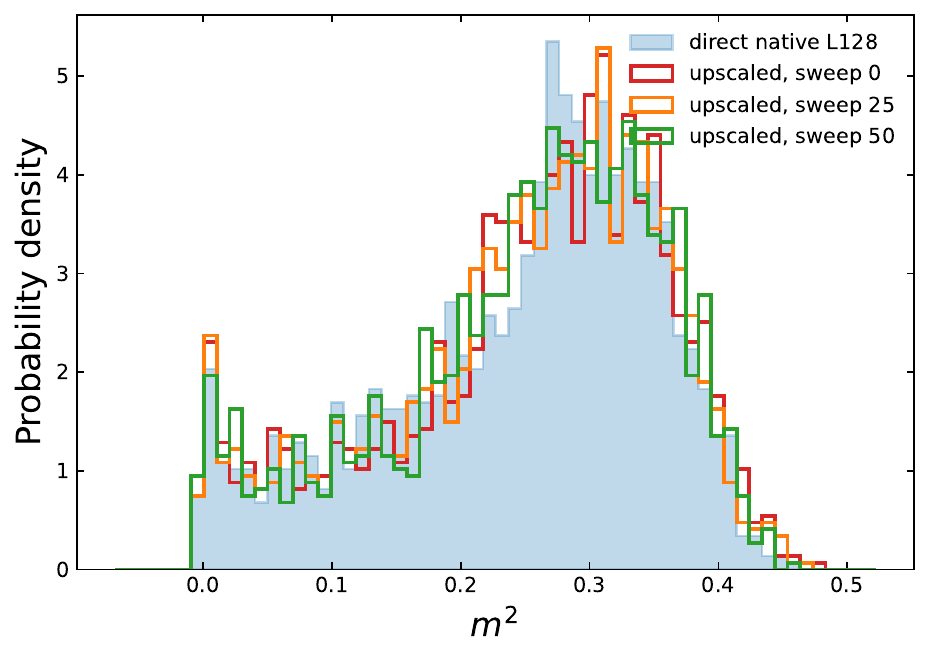}
      \includegraphics[width=0.3\linewidth]{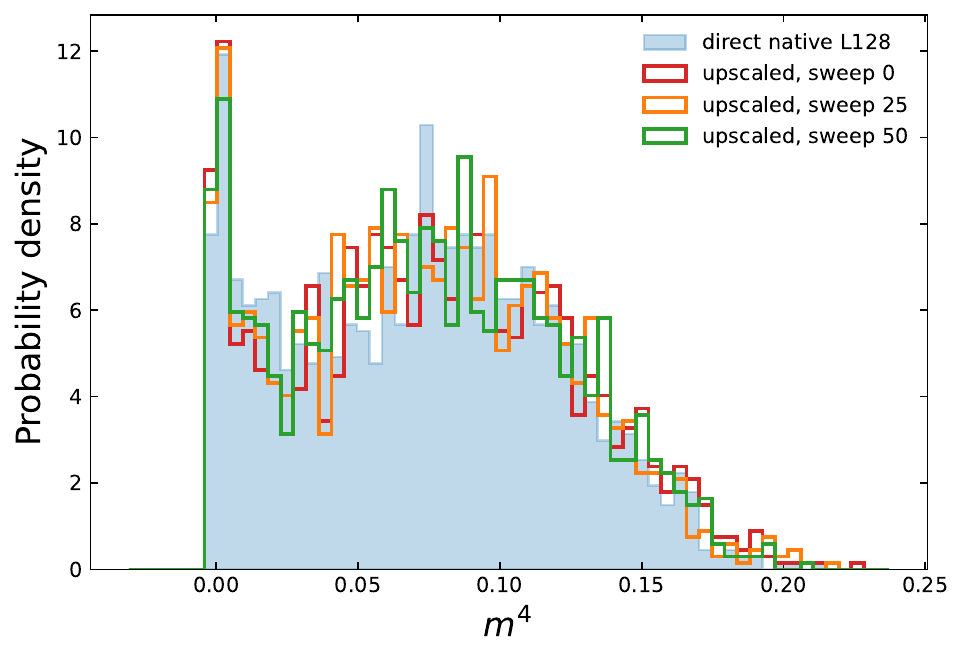}
      \includegraphics[width=0.3\linewidth]{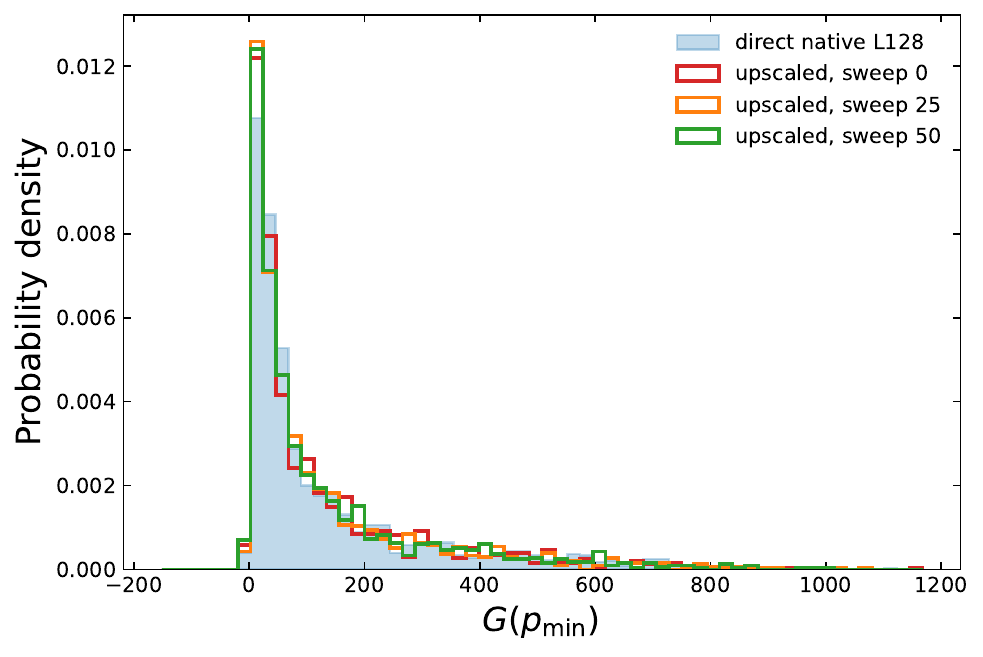}
      \caption{Rethermalization of the $L=64\to128$ upscaled ensemble.  The panels show
the same observables as Fig.~\ref{fig:L32-to-64-rethermalization}.
}
      \label{fig:L64-to-128-rethermalization}
\end{figure*}

\clearpage
\bibliography{rg_guided_nf_refs}

\end{document}